\RequirePackage{fix-cm}

\documentclass[smallextended]{svjour3}       % onecolumn (second format)
\smartqed  % flush right qed marks, e.g. at end of proof
\usepackage{graphicx}
\usepackage{amsmath}
\usepackage{amssymb}
\usepackage{booktabs}
\usepackage{algorithm}
\usepackage{algorithmic}
\usepackage{multirow}
\usepackage{tabularx}
\usepackage{xurl}
\usepackage{natbib}
\usepackage[most]{tcolorbox}
\usepackage{tikz}
\usetikzlibrary{positioning,arrows.meta,fit,backgrounds,calc}
\newtcolorbox{summarybox}{breakable,colback=gray!5,colframe=gray!60,boxrule=0.5pt,arc=2pt,left=6pt,right=6pt,top=4pt,bottom=4pt,fonttitle=\bfseries,title=Summary}

\begin{document}

\title{What Should a Software Team Spend on Agents? Simulation-Based Cost Optimization of Mixed AI-Human Teams}
\titlerunning{Simulation-Based Cost Optimization of Mixed AI-Human Teams}

\author{Steffen Herbold}

%\authorrunning{Herbold}

\institute{Steffen Herbold, 0000-0001-9765-2803 \at
University of Passau, Passau, Germany.
\email{steffen.herbold@uni-passau.de}
}

\date{Received: date / Accepted: date}

\maketitle

\begin{abstract}
\textbf{Context:} AI agents are rapidly adopted in software development, but their token costs are rising and it is unclear how much a team should spend on agents, and on which tasks.\\
\textbf{Objective:} We determine how teams should divide spending between human developers and AI agents to deliver the features a project demands at the lowest total cost.\\
\textbf{Method:} We model mixed teams of human developers and AI agents with eight specialized roles as a discrete-time stochastic process covering bug injection and detection, technical debt, human supervision of agents, and release cycles with feature freezes. Using Monte Carlo simulation and evolutionary search, we determine cost-optimal teams for 432 scenarios varying developer capability, project velocity, debt dynamics, agent capability and price, and bug costs, and compare them with a human-only baseline.\\
\textbf{Results:} Escaped bugs are the dominant cost driver, and 61\% of the agent budget goes to finding bugs. Agent capability matters more than token price, since higher-quality outputs reduce downstream quality assurance costs. In teams with capable agents, human developers mostly fix bugs and leave the implementation of new features to the agents. Teams with high-capability agents are typically cheaper than human-only teams, but low-capability agents offer no consistent advantage. At realistic velocities, the optimal teams still spend more on developers than on agents.\\
\textbf{Conclusions:} There is neither a universal token budget per developer nor a universal agentic team structure. As in traditional development, this depends on many factors and teams need to be carefully tailored to the project context to achieve the optimal cost-quality trade-off. 
\end{abstract}


% ============================================================
\section{Introduction}
\label{sec:introduction}

\begin{quote}
  \textit{If that \$500{,}000 engineer did not consume at least \$250{,}000 worth of tokens, I am going to be deeply alarmed} --- Jensen Huang, CEO of Nvidia (March 2026).
\end{quote}

Agentic AI systems based on Large Language Models (LLMs), hereafter referred to as agents, have been shown to have strong capabilities in software engineering tasks, including code generation~\citep{aleithan2024swebenchplus,merrill2026terminal}, code review~\citep{Hu2026}, and natural language focused software engineering activities~\citep{pena2026selusoftwareengineeringlanguage}. This has led to a rapid adoption of these agents in software development teams~\citep{bogdanov2026agentadoption}. Recently, the uncritical adoption and unrestricted use of these agents is being rolled back~\citep{juricic2026metaclaude}, not least due to the spiraling costs. Such problems are not unexpected with new technology, where we are still trying to understand all implications of AI-native software development~\citep{hassan2026ainative}. This raises the question whether Jensen Huang's general estimate that developers should consume at least half of their salary in agent tokens is reasonable, or whether there are better ways to compose human and agent workforces. This is the central research question of this work:

\begin{description}
  \item[\textbf{RQ}] How should software development teams compose their human and agent workforce to deliver the work a project demands at the lowest total cost, counting the cost of the developers, the cost of the agents, and the cost of the bugs that reach production?
\end{description}

To answer this question, we develop a simulation model of mixed AI-human software development teams. The core assumption of our model is that the human developers still own the code, take responsibility for the quality, and guide the key design decisions. Agents support the human developers by implementing features, fixing bugs, and providing quality assurance. However, the agents also drain resources from the human developers because they require supervision, need feedback on design decisions, or help resolve issues where they fail autonomously. 

We simulate the development of a software product represented as a stochastic process that evolves over multiple sprints, where each sprint represents a fixed-duration development iteration. Within each sprint, human developers and agents implement features, inadvertently introduce bugs, as well as fix known bugs. Bugs can be detected by code reviews, more expensive audits for security issues, and generated tests. Implementing features can also require changes to the architecture, which are more expensive to implement and require the sign-off by the human developers. Neglecting architecture maintenance, code documentation, and refactoring accumulates debt, which can be reduced by agents investing effort in these activities to prevent a slow-down of the development. 

Within our model, we distinguish between eight different agent roles for different development and quality assurance tasks, so that we can not only analyze how to spend on agents but also how to allocate this money for different purposes. Costs are estimated based on the estimated token use by agents, salaries of human developers, and the cost of bugs that reach production. Using a simple evolutionary search, we determine optimal team compositions for different scenarios, e.g., different assumptions on agent skills and costs, development speed, and human capabilities and costs. 

Our simulations show that the costs of agentic software development and the optimal team compositions depend on many interacting factors. The key findings are:
\begin{itemize}
  \item \emph{Agent capability matters more than agent cost.} High-capability agents require less human supervision, reducing personnel costs, while also creating fewer bugs, which reduces the number of bugs that escape to production. Consequently, differences in costs for tokens matter less than the capability differences. 
  \item \emph{Capable agents turn developers into bug fixers}. For the most capable and cheap agent tiers we tested, developers should spend a median of 88--90\% of their coding time on bug fixing, while at the same time, few agents are used to fix bugs. 
  \item \emph{Agents only pay off if they are good enough.} Under weaker capability assumptions, agent-supported development is not consistently cheaper than our human-only baseline. More optimistic estimates of agent capability nevertheless show that significant cost savings are possible. 
  \item \emph{Agents enable velocities that are impossible to achieve with human-only teams.} The throughput of agents can be scaled to extreme development speeds, possibly enabling new classes of products and more rapid development than can be achieved by human developers. 
  \item \emph{In most of our scenarios, costs for human developers exceed those for agents.} Unless extreme or high velocities force a strong reliance on agents, our simulation still favors paying more for human developers than for agents. 
  \item \emph{Escaped bugs, not labor, are the main cost driver.} Because escaped bugs are expensive, the optimal teams invest heavily in quality assurance. Even so, escaped (security) bugs remain the largest cost factor.
  \item \emph{Quality assurance consumes most of the agent budget.} Since preventing bugs saves money, quality assurance activities are often scaled to high effort and therefore dominate the costs for agents. In other words, getting features is cheap, but preserving quality is expensive.
\end{itemize}

% ============================================================
\section{Related work}
\label{sec:related}
% ============================================================

To the best of our knowledge, there is no directly related work, i.e., attempts to determine optimal human-agent team structures accounting for costs and quality of a software product, either through simulation or through other methods. Note that the quantification of capabilities of human developers or AI agents is not discussed here, as this is not the objective of this work. Instead, we discuss the relevant work that we use to estimate the dynamics of our simulation model in Section~\ref{sec:setup}. Further, our work does not discuss how exactly humans and AI work together or how processes are defined. Instead, we refer to the emerging literature on AI-based software engineering in this regard, e.g., \cite{hassan2026ainative, hassan2025agentic, li2025aiteammates}. We therefore focus on earlier methods for simulating software projects, which inspired this work.

Simulation is a common experimental method to explore how situations evolve, given a model of the underlying system --- in our case the dynamics of software development. \cite{abdelhamid1991dynamics} built the first such model for system dynamics and showed that a simulation based on staffing, schedule pressure, productivity, and defect generation can reproduce the behavior of a real-world project. \cite{kellner1999simulation} emphasize that the simulation models serve a specific purpose. For software engineering, discrete-event simulations are the most common approach \citep{zhang2008reflections,garciagarcia2020simulation}, since software is typically considered to be made up of artifacts that are evolved through time-discrete changes instead of through a continuous model, e.g., to simulate agile development~\citep{cao2010agile}. 

Agent-based simulation methods use a similar modeling approach to us. In such methods, agents act on and observe a simulation environment, typically with the goal to achieve some reward. The simulation then defines the types of agents and their behavior, e.g., to model human software developers as agents~\citep{smith2006agentbased}. We note that this usage of the term \textit{agent} differs from the AI agents for software development that we consider here. To avoid confusion, we hereafter use the term actor-based simulation and speak of actors when referring to aspects of agent-based simulation. 
Actor-based simulation has been further explored in the literature, e.g., to simulate developer decisions~\citep{agarwal2010flexible}, or to estimate effort and time required for projects~\citep{spasic2012avl}. \cite{honsel2015mining} suggest using repository mining to model the behavior of human actors, and demonstrate that parameters estimated with repository mining can lead to accurate simulations of the evolution of open-source projects~\citep{honsel2021investigation}. 

The simulation model we develop is similar to the above, i.e., we have distinct actors with different behavior that drive the evolution of a software system through time-discrete steps and we estimate the behavior of human actors using results from repository mining research and reports on developer productivity. However, none of the above takes AI agents into account, so none of these models is directly comparable to our work. This is the gap that our research addresses, i.e., a simulation framework that allows us to model how human developers and AI agents co-evolve software.

% ============================================================
\section{Model description}
\label{sec:model}
% ============================================================

We now describe our simulation model in which we treat software development as a discrete-time stochastic dynamical system, in line with prior work on software simulation (see Section~\ref{sec:related}). The simulation runs for $T$ sprints, where each sprint represents one
fixed-duration iteration. States are indexed $t \in \{0, 1, \ldots, T\}$,
where $S(0)$ is the fixed initial state (before any sprint executes) and
$S(t)$ for $t \geq 1$ is the state produced by sprint $t$. The simulation configuration is defined by the team composition and process parameters 
$\mathbf{c} = (N, \mathbf{A}_\mathrm{dev}, \mathbf{A}_\mathrm{qa}, R, F_\mathrm{freeze}, f_\mathrm{fix})$. The parameters
$N$, $f_\mathrm{fix}$, $\mathbf{A}_\mathrm{dev}$, $\mathbf{A}_\mathrm{qa}$ define the team structure (Sections~\ref{sec:model:roles} and~\ref{sec:model:team}). The parameters $R$ and $F_\mathrm{freeze}$ define the release cycle and feature freezes (Section~\ref{sec:model:cycle}). The team composition further depends on whether the current sprint is in the development phase or the feature freeze phase of the release cycle (Section~\ref{sec:model:team}). The development team of humans and agents evolves a software project whose state at sprint $t$ is represented by a 22-dimensional state vector $S(t)$ defined in Section~\ref{sec:model:state}. Scenario parameters govern the behavior of the underlying software development process and the capabilities of the agents, such as feature throughput rates, bug injection rates, bug detection probabilities, and cost coefficients (Section~\ref{sec:model:parameters}). The transition from $S(t-1)$ to $S(t)$ is governed by the sprint dynamics $\mathcal{T}$, which consists of ten ordered steps (Section~\ref{sec:model:transition}).

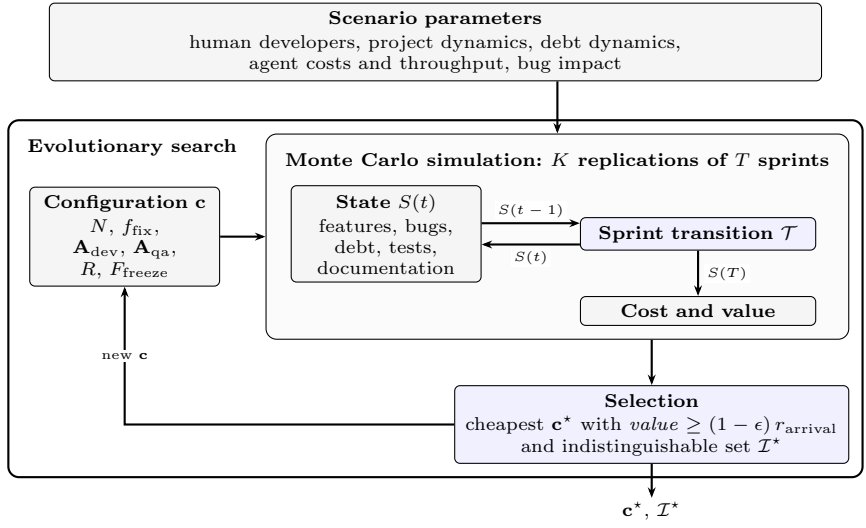
\begin{figure}[t]
\centering
% Overview of the model structure (Section 3), included from main.tex.
\pgfdeclarelayer{back}
\pgfsetlayers{back,background,main}
\begin{tikzpicture}[
    font=\scriptsize,
    >={Stealth[length=4pt]},
    box/.style={draw, rounded corners=2pt, fill=gray!8, align=center,
                inner sep=3pt},
    proc/.style={box, fill=blue!6},
    frame/.style={draw, rounded corners=4pt, inner sep=5pt},
    title/.style={font=\scriptsize\bfseries, anchor=north west, inner sep=2pt},
    flow/.style={->, thick},
    lbl/.style={font=\tiny, fill=white, inner sep=1pt},
  ]

  % ---- simulation ----
  \node[box, text width=2.3cm] (state) at (0,0) {%
    \textbf{State $S(t)$}\\[1pt]
    features, bugs, debt, tests, documentation};
  \node[proc, text width=2.9cm, right=1.3cm of state] (trans) {%
    \textbf{Sprint transition $\mathcal{T}$}};
  \draw[flow] ([yshift=4pt]state.east) -- node[lbl, above=1pt] {$S(t-1)$} ([yshift=4pt]trans.west);
  \draw[flow] ([yshift=-4pt]trans.west) -- node[lbl, below=1pt] {$S(t)$} ([yshift=-4pt]state.east);
  \node[box, text width=2.9cm, below=0.6cm of trans] (obj) {%
    \textbf{Cost and value}};
  \draw[flow] (trans) -- node[lbl, right=2pt] {$S(T)$} (obj);

  \node[title] (mctitle) at ($(state.north west)+(-0.15,0.5)$)
    {Monte Carlo simulation: $K$ replications of $T$ sprints};
  \begin{pgfonlayer}{back}
    \node[frame, fill=gray!3, fit=(mctitle)(state)(trans)(obj)] (mc) {};
  \end{pgfonlayer}

  % ---- configuration, left of the simulation ----
  \node[box, text width=2.3cm, left=0.6cm of mc] (config) {%
    \textbf{Configuration $\mathbf{c}$}\\[1pt]
    $N$, $f_\mathrm{fix}$,\\ $\mathbf{A}_\mathrm{dev}$, $\mathbf{A}_\mathrm{qa}$,\\ $R$, $F_\mathrm{freeze}$};
  \draw[flow] (config) -- (mc.west |- config);

  % ---- selection ----
  \node[proc, text width=5.0cm, anchor=north east] (select) at ($(mc.south east)+(0,-0.6)$) {%
    \textbf{Selection}\\[1pt]
    cheapest $\mathbf{c}^\star$ with $\mathit{value} \geq (1-\epsilon)\,r_\mathrm{arrival}$\\
    and indistinguishable set $\mathcal{I}^\star$};
  \draw[flow] (mc.south -| select.north) -- (select.north);

  % ---- feedback: new candidates ----
  \draw[flow] (select.west) -| node[lbl, pos=0.75] {new $\mathbf{c}$} (config.south);

  \node[title] (searchtitle) at ($(config.west |- mc.north)+(-0.1,0)$)
    {Evolutionary search};
  \begin{pgfonlayer}{back}
    \node[frame, thick, fit=(searchtitle)(config)(mc)(select)] (search) {};
  \end{pgfonlayer}

  % ---- scenario parameters, fixed outside the search ----
  \node[box, text width=10.0cm, anchor=south] (params) at ($(search.north)+(0,0.5)$) {%
    \textbf{Scenario parameters}\\[1pt]
    human developers, project dynamics, debt dynamics,\\ agent costs and throughput, bug impact};
  \draw[flow] (params.south -| mc.north) -- (mc.north);

  % ---- output ----
  \draw[flow] (select.south) -- ++(0,-0.45) node[below, inner sep=1pt] {$\mathbf{c}^\star$, $\mathcal{I}^\star$};
\end{tikzpicture}
\caption{Overview of the model. The evolutionary search proposes configurations $\mathbf{c}$, consisting of the human staffing level $N$, the share $f_\mathrm{fix}$ of the developers' coding time spent on bug fixing, the agent compositions $\mathbf{A}_\mathrm{dev}$ and $\mathbf{A}_\mathrm{qa}$ for development and feature-freeze sprints, the number of sprints per release $R$, and the feature-freeze length $F_\mathrm{freeze}$. Each feasible configuration is evaluated with $K$ Monte Carlo replications of a simulation that applies the sprint transition $\mathcal{T}$ to the state $S(t-1)$ to obtain $S(t)$ for $T$ sprints. The costs and the $\mathit{value}$, i.e., the features completed per sprint, are computed from the terminal state $S(T)$. The selection determines the cheapest configuration $\mathbf{c}^\star$ whose value reaches the feature arrival rate $r_\mathrm{arrival}$ up to a tolerance $\epsilon$, and the set $\mathcal{I}^\star$ of configurations that are statistically indistinguishable from it.}
\label{fig:overview}
\end{figure}

We derive the metrics that describe how well a team configuration performs from the terminal state and the per-sprint record of the work the agents completed (Section~\ref{sec:model:objectives}): the $value$ as the average number of features completed per sprint, and three cost drivers, namely the cost caused by the bugs that escape to production, the cost of the human developers, and the cost of the agents. We formulate a search problem that minimizes the costs, subject to the constraint that a configuration keeps up with the work the project generates (Section~\ref{sec:model:sosp}). By running multiple simulations with different random seeds for each configuration $\mathbf{c}$, we can establish the uncertainty of the simulation through a Monte Carlo approach and estimate the confidence intervals of the costs (Section~\ref{sec:model:mc}). Since the estimates are noisy, we do not only report a single best configuration, but also the set of configurations that cannot be statistically distinguished from it (Section~\ref{sec:model:selection}). Since the search space is combinatorial and high-dimensional, we further describe the search strategy we use to efficiently explore the space of valid configurations (Section~\ref{sec:model:search}). Figure~\ref{fig:overview} summarizes how these parts fit together.

% ---- 3.1 ----
\subsection{Roles}
\label{sec:model:roles}

Similar to \cite{hassan2026ainative}, we assume that agents will conduct most of the actual software development work. Therefore, there is only a single human role included in our model: the \textit{Human
Developer}. Developers communicate with each other, supervise agents,
review and approve architecture proposals, resolve
escalations that agents cannot handle autonomously, and contribute new
features and bug fixes with the remaining time.

We do not assume that developers dedicate all of their time to the project. $N$ is therefore
not a headcount but a \emph{staffing level} measured in full-time equivalents (FTE), and it does
not need to be an integer: $N = 0.25$ describes one developer who devotes a quarter of their
working time to the project, and $N = 2.5$ describes two full-time developers plus one
half-time developer. Accordingly, $H_\mathrm{capacity}$ is the gross time budget of one FTE
per sprint, so that the gross budget available to the whole team is $N \cdot
H_\mathrm{capacity}$ hours per sprint.

We have eight specialist AI agent roles, indexed $k = 1,\ldots,8$. All agents of the
same type are assumed to be statistically identical. We further distinguish between agents that
require supervision and agents that run fully autonomously. The supervised agents consume monitoring capacity from
the human developers but are otherwise assumed to execute tasks autonomously. Table~\ref{tab:roles} summarizes each AI agent role's primary function. We note that different roles do not mean that there must be different AI systems or agentic frameworks. This separation is rather introduced to enable the simulation to have distinct estimates for the agentic costs and capabilities for different aspects of software development and also to allow the allocation of these costs to optimize quality while minimizing costs. 

The eight roles fall into three functional groups. The production roles
($k = 1,\ldots,4$) all directly contribute to the production code. 
The \textit{Feature Implementation Agent} is the primary
productivity driver implementing features and architecture changes. While doing so, the agent also generates
architecture change requests and introduces new bugs into the codebase.
The \textit{Architecture Agent} creates proposals for architecture updates for human review. Neglecting architecture change requests degrades internal quality and will lead to architectural debt accumulation.
The \textit{Bug Fixing Agent} drains the known-bug
queue. The work conducted by the Feature Implementation and Bug Fixing Agents introduces technical debt.
The \textit{Refactoring Agent} improves the internal quality of the system autonomously and reduces technical debt.
From a conceptual perspective, the \textit{Architecture Agent} is responsible for high-stakes structural changes that require human review, while the \textit{Refactoring Agent} is responsible for low-stakes internal improvements that can be executed autonomously without review. Consequently, the Refactoring Agent is not supervised and consumes no monitoring capacity from the human developers. 

The two review services ($k = 5,6$) inspect
every code change synchronously in the same sprint as injection. The \textit{Code Review
Agent} detects all kinds of bugs but gets less effective if there is more code to review. The \textit{Security Audit Agent} can only detect security bugs and allows the modeling of more expensive agents for this high-stakes role. Like the Code Review Agent, it becomes less effective with larger review volumes and more technical debt. These two agents run autonomously, i.e., the human developers do not need to supervise them and detected bugs are automatically added to the bug queues.
The \textit{Test Development Agent} ($k = 7$) builds up a continuously growing test suite, which helps detect bugs.

Finally, there is one \emph{support role} ($k = 8$): the \textit{Developer Documentation Agent} creates documentation to aid human developers, which in turn reduces the
per-agent monitoring overhead that consumes human capacity by reducing the cognitive load
of the human developers when supervising agents.

\begin{table}[t]
\caption{The eight AI agent roles.}
\label{tab:roles}
\footnotesize
\begin{tabularx}{\linewidth}{cp{2.6cm}X}
\toprule
$k$ & Role & Primary Function \\
\midrule
1 & Feature Implementation
  & Implements features and approved architecture changes \\[2pt]
2 & Architecture \& Design
  & Creates architecture and design proposals for human review \\[2pt]
3 & Bug Fixing
  & Diagnoses and repairs known bugs \\[2pt]
4 & Refactoring
  & Restructures code to reduce the technical debt \\[2pt]
5 & Code Review
  & Reviews all code changes to detect bugs \\[2pt]
6 & Security Audit
  & Targeted review to identify security-class vulnerabilities \\[2pt]
7 & Test Development
  & Writes automated tests and grows test coverage for a continuous integration system to detect bugs \\[2pt]
8 & Developer Documentation
  & Produces documentation (e.g., API references, architecture guides, and contributor docs) to aid human developers \\
\bottomrule
\end{tabularx}
\end{table}

% ---- 3.2 ----
\subsection{Release cycle and feature freezes}
\label{sec:model:cycle}

A release occurs exactly at sprint indices $R, 2R, 3R, \ldots$ where $R \in \mathbb{N}_{\geq 1}$
is the fixed release cadence. At each release, unfixed bugs (both hidden and known)
escape to production, incrementing the cumulative escaped bug counters. 
As in many real-world projects, we model a feature freeze to stabilize the
codebase before release. The feature freeze lasts for $F_\mathrm{freeze} \in
\{0, \ldots, R-1\}$ sprints before a release.
Sprint $t$ is in the freeze window iff $((t-1) \bmod R) \geq R - F_\mathrm{freeze}$, i.e., the freeze window comprises the last $F_\mathrm{freeze}$ sprints of each release cycle, including the release sprint itself.

Each sprint uses one of two agent compositions, depending on its position in the release cycle. The \emph{development composition} is used during the first $R - F_\mathrm{freeze}$ sprints of each cycle, and the \emph{QA composition} during the remaining $F_\mathrm{freeze}$ sprints. The different compositions allow a different focus,
i.e., focus on feature addition versus focus on quality assurance and bug fixing.

% ---- 3.3 ----
\subsection{Project and team structure}
\label{sec:model:team}

Given the above description of the roles and release cycle, we can now specify the project and team structure using six control variables: the human staffing level in FTE
$N \in \mathbb{R}_{>0}$, the number of sprints per release $R \in \mathbb{N}_{\geq 1}$, the length of the feature freeze period $F_\mathrm{freeze} \in
\{0, \ldots, R-1\}$, the share $f_\mathrm{fix} \in [0,1]$ of the developers' coding time that goes to bug fixing rather than feature implementation, and two 8-element agent composition vectors $\mathbf{A}_\mathrm{dev}$ and
$\mathbf{A}_\mathrm{qa}$ described below.

$\mathbf{A}_\mathrm{dev} = (a_{dev,1},\ldots,a_{dev,8}) \in \mathbb{N}_{\geq 0}^{8}$ is the
development-phase composition and $\mathbf{A}_\mathrm{qa} = (a_{qa,1},\ldots,a_{qa,8}) \in \mathbb{N}_{\geq 0}^{8}$ is the QA composition. As established in Section~\ref{sec:model:roles}, the roles fall into two groups with respect to human oversight: the \emph{supervised} roles
$\mathcal{K}_\mathrm{sup} = \{1,2,3,7,8\}$ and the autonomous roles
$\mathcal{K}_\mathrm{aut} = \{4,5,6\}$, i.e., the Refactoring Agent and the two review
services, whose work the developers do not review (Section~\ref{sec:model:roles}). We write
\begin{equation}
  n_\mathrm{sup}(t) = \sum_{k \in \mathcal{K}_\mathrm{sup}} a_{\cdot,k}
  \label{eq:nsup}
\end{equation}
for the number of active supervised agents in sprint $t$, where the placeholder $\cdot$ stands for the composition that is active in sprint $t$, i.e., $\mathrm{dev}$ during development sprints and $\mathrm{qa}$ during feature-freeze sprints. We use this notation throughout wherever a statement holds for both phases. Every supervised agent consumes monitoring effort from the human developers, so the admissible compositions are limited by the human-capacity feasibility condition of Section~\ref{sec:model:transition}, Eq.~\eqref{eq:feas_monitoring}. Leaving supervised slots unfilled may be optimal in some cases to reduce coordination overhead and free up human capacity for review, escalation, and implementation work.
None of the autonomous roles consumes human monitoring capacity. For the two review services, the counts $a_{\cdot,5}$ and $a_{\cdot,6}$ are therefore unbounded and only limited by their token cost, since at some point investing more in detecting possible bugs yields diminishing returns. The Refactoring Agent is unsupervised in the same way, but the debt it can remove is limited by the debt the team accumulates, so its count stays bounded by the work that reaches it. The dev-phase and QA-phase counts are independent of each other, so a service can be staffed differently during a feature freeze than during development.
$\mathbf{A}_\mathrm{dev}$ must satisfy $a_{dev,1} \geq 1$, i.e., there must be at least one Feature
Implementation Agent in the development phase. This also ensures that there is $value$, i.e., that at least some features are implemented and the simulation does not trivially collapse to zero throughput.
For $F_\mathrm{freeze} \geq 1$, $\mathbf{A}_\mathrm{qa}$ is required to have $a_{qa,1} = a_{qa,2} = 0$, to ensure that no new features are added during a feature freeze and that all agents are redeployed to QA roles. If $F_\mathrm{freeze}=0$, the QA composition is never active and has no effect on the simulation. We therefore fix $\mathbf{A}_\mathrm{qa} = \mathbf{0}$ in that case.

% ---- 3.4 ----
\subsection{State space}
\label{sec:model:state}

$S(t)$ denotes the 22-component system state produced by sprint $t$ (for $t \geq 1$), and $S(0)$ denotes the fixed initial state. The components fall into four
clusters summarized in Table~\ref{tab:state}: three feature variables, twelve bug variables, five variables for technical debt and the human capacity queues, and two variables for test infrastructure and documentation coverage.

\subsubsection{Feature pipeline}
$F_\mathrm{backlog}(t) \in \mathbb{N}_{\geq 0}$ counts features awaiting
implementation after sprint $t$ completed. Within each sprint, this queue grows according to the feature arrival rate (see Section~\ref{sec:model:parameters}), while the \textit{Feature Implementation Agents} and the
direct coding capacity of \textit{Human Developers} reduce it by implementing features and thereby growing the monotonically increasing cumulative count of fully implemented features $F_\mathrm{completed}(t) \in \mathbb{N}_{\geq 0}$. $F_\mathrm{cycle}(t) \in \mathbb{N}_{\geq 0}$ counts the features implemented since the last release.

\subsubsection{Bug queues}

Our simulation model distinguishes between three mutually exclusive bug types: \emph{minor}, \emph{major}, and \emph{security} bugs. Configuration parameters determine the impact of the different bug types on quality and the rates at which bugs of the different types are introduced, relative to each other (see Section~\ref{sec:model:parameters}). The bugs are tracked in four kinds of queues (hidden, known, released, and escaped) with one counter per bug type, giving twelve bug state variables in total.

$B_\mathrm{hid,x}(t)$ (for $x \in \{\mathrm{min, maj, sec}\}$) counts type-$x$ bugs
that have been injected in the current release cycle but have not been detected yet.
$B_\mathrm{kno,x}(t)$ counts type-$x$ bugs that have been detected and are queued for
repair. This separation is important, because only detected bugs can be fixed, while hidden bugs remain unnoticed in the code until they are detected or escape to production. $B_\mathrm{esc,x}(t) \in \mathbb{N}_{\geq 0}$ is the
monotonically increasing cumulative count of type-$x$ bugs that reached production.
Since escaped bugs must still be fixed, $B_\mathrm{rel,x}(t)$ counts the subset of
escaped type-$x$ bugs still open after sprint $t$. The Bug Fixing Agent fixes these bugs before those in the known-bug queue.

\subsubsection{Technical debt and human capacity queues}

$D_\mathrm{tech}(t) \in [0, 1]$ is a normalized measure of technical debt: $D_\mathrm{tech} = 0$ represents
a debt-free codebase and $D_\mathrm{tech} = 1$ represents maximum degradation. $D_\mathrm{tech}(t)$ can be reduced autonomously by the \textit{Refactoring Agent}. $Q_\mathrm{arch}(t) \in \mathbb{N}_{\geq 0}$ counts the architecture change requests that are still pending, $Q_\mathrm{rev}(t) \in \mathbb{N}_{\geq 0}$ counts the architecture proposals awaiting human review, and $Q_\mathrm{arch,impl}(t) \in \mathbb{N}_{\geq 0}$ counts the approved architecture proposals that are awaiting implementation. Together, these three variables measure the architectural degradation, which requires human intervention to be reduced and cannot be reduced fully autonomously. $Q_\mathrm{esc}$ is the agent escalation queue: agents may fail to implement a feature on their own and need to escalate this to a human developer to resolve the issue. 

All these variables grow with the work of the Feature Implementation Agents and the Bug Fixing Agents, which introduce technical debt and generate architecture change requests as they implement features and fix bugs. Thus, faster development speed directly increases the accumulation of technical debt, architectural debt, and escalation requests.

\subsubsection{Test infrastructure and documentation}

We assume that there is a continuous integration system in place that automatically runs a test suite to detect
bugs. The status of this test suite is tracked by $N_\mathrm{test}(t)$, which is the cumulative number of tests written by the Test
Development Agent. New tests are added every sprint to increase the \emph{test coverage} $\mathrm{cov\_test}$, which is derived from the test count and the number of implemented features as
\begin{equation}
  \mathrm{cov\_test}(t) = \min\!\left(c_\mathrm{test,ceil},\;
    \frac{N_\mathrm{test}(t)}{\max\!\bigl(1,\,F_\mathrm{completed}(t)\bigr)}\right).
  \label{eq:coverage}
\end{equation}
Higher coverage increases the likelihood that introduced bugs are detected and move from the hidden to the known queues. 

Similarly, $N_\mathrm{doc}(t)$ is the cumulative number of documentation items produced by
the Developer Documentation Agent and grows monotonically each sprint.
Documentation \emph{coverage} is derived from this count in the same way as test coverage,
but relative to $F_\mathrm{completed}(t-1)$:
\begin{equation}
  \mathrm{cov\_doc}(t-1) = \min\!\left(1,\;
    \frac{N_\mathrm{doc}(t-1)}{\max\!\bigl(1,\,F_\mathrm{completed}(t-1)\bigr)}\right),
  \label{eq:doc_cov}
\end{equation}
capped at 1, i.e., full documentation coverage.
The $t-1$ index reflects that the documentation is updated at the end of the sprint, i.e., the new documentation only becomes available in the next sprint. This means that the benefits of documentation investment (i.e., reduced monitoring effort for human developers) are always lagged by one sprint.

\begin{table}[t]
\caption{State vector components with domains and the sprint-transition
  steps (cf.\ Section~\ref{sec:model:transition}) that write each variable.}
\label{tab:state}
\footnotesize
\begin{tabularx}{\linewidth}{p{2.3cm}Xp{1.5cm}p{1.4cm}}
\toprule
Symbol & Description & Domain & Modified (step) \\
\midrule
$F_\mathrm{backlog}(t)$     & Features awaiting implementation                      & $\mathbb{N}_{\geq 0}$  & 3 \\[2pt]
$F_\mathrm{completed}(t)$   & Cumulative features shipped            & $\mathbb{N}_{\geq 0}$  & 3 \\[2pt]
$F_\mathrm{cycle}(t)$       & Features implemented in the current release cycle & $\mathbb{N}_{\geq 0}$  & 3, 10 \\[2pt]
\midrule
$B_\mathrm{hid,min}(t)$     & Hidden minor bugs in the current release cycle           & $\mathbb{N}_{\geq 0}$  & 5, 10 \\[2pt]
$B_\mathrm{hid,maj}(t)$     & Hidden major bugs in the current release cycle          & $\mathbb{N}_{\geq 0}$  & 5, 10 \\[2pt]
$B_\mathrm{hid,sec}(t)$     & Hidden security bugs in the current release cycle        & $\mathbb{N}_{\geq 0}$  & 5, 10 \\[2pt]
$B_\mathrm{kno,min}(t)$     & Known minor bugs queued for repair in the current release cycle & $\mathbb{N}_{\geq 0}$  & 5, 7, 10 \\[2pt]
$B_\mathrm{kno,maj}(t)$     & Known major bugs queued for repair in the current release cycle & $\mathbb{N}_{\geq 0}$  & 5, 7, 10 \\[2pt]
$B_\mathrm{kno,sec}(t)$     & Known security bugs queued for repair in the current release cycle & $\mathbb{N}_{\geq 0}$ & 5, 7, 10 \\[2pt]
$B_\mathrm{rel,min}(t)$     & Open in-production minor bugs requiring repair              & $\mathbb{N}_{\geq 0}$  & 7, 10 \\[2pt]
$B_\mathrm{rel,maj}(t)$     & Open in-production major bugs requiring repair              & $\mathbb{N}_{\geq 0}$  & 7, 10 \\[2pt]
$B_\mathrm{rel,sec}(t)$     & Open in-production security bugs requiring repair           & $\mathbb{N}_{\geq 0}$  & 7, 10 \\[2pt]
$B_\mathrm{esc,min}(t)$     & Cumulative minor bugs escaped              & $\mathbb{N}_{\geq 0}$  & 10 \\[2pt]
$B_\mathrm{esc,maj}(t)$     & Cumulative major bugs escaped               & $\mathbb{N}_{\geq 0}$  & 10 \\[2pt]
$B_\mathrm{esc,sec}(t)$     & Cumulative security bugs escaped            & $\mathbb{N}_{\geq 0}$  & 10 \\[2pt]
\midrule
$D_\mathrm{tech}(t)$                      & Technical debt level                                   & $[0, 1]$               & 9 \\[2pt]
$Q_\mathrm{rev}(t)$         & Architecture review queue length                                                              & $\mathbb{N}_{\geq 0}$  & 6 \\[2pt]
$Q_\mathrm{arch}(t)$        & Architecture change-request queue awaiting agent resolution  & $\mathbb{N}_{\geq 0}$  & 6 \\[2pt]
$Q_\mathrm{arch,impl}(t)$   & Approved architecture proposals awaiting implementation    & $\mathbb{N}_{\geq 0}$  & 6 \\[2pt]
$Q_\mathrm{esc}(t)$         & Escalation queue length                     & $\mathbb{N}_{\geq 0}$  & 8 \\[2pt]
\midrule
$N_\mathrm{test}(t)$        & Cumulative tests written     & $\mathbb{R}_{\geq 0}$  & 4 \\[2pt]
$N_\mathrm{doc}(t)$         & Cumulative documentation items written & $\mathbb{R}_{\geq 0}$  & 9 \\
\bottomrule
\end{tabularx}
\end{table}

% ---- 3.5 ----
\subsection{Configuration parameters}
\label{sec:model:parameters}

The behavior of the agents and the simulation model in each sprint
is governed by a set of parameters that must be defined before running the simulation. 
The parameters codify the assumptions about the underlying software development process and the capabilities of the agents, and they determine how the state variables evolve over time. They can be grouped into human-capacity parameters (Table~\ref{tab:params:human}), feature-pipeline and throughput-rate parameters (Table~\ref{tab:params:pipeline}), bug-injection parameters (Table~\ref{tab:params:injection}), quality assurance parameters for bug detection (Table~\ref{tab:params:detection}), technical-debt parameters (Table~\ref{tab:params:debt}), and cost parameters (Table~\ref{tab:params:costs}). The tables describe the meaning of each parameter, its domain, and which step of the sprint transition first uses it (see Section~\ref{sec:model:transition} for the step definitions). Section~\ref{sec:model:transition} describes how each parameter is used, and Section~\ref{sec:setup} describes how we set their values. The human-only baseline (Section~\ref{sec:setup:baseline}) requires two more human parameters and one more decision variable that are not part of the model described here. We introduce them together with the baseline.

\begin{table}[t]
\caption{Human-capacity parameters.}
\label{tab:params:human}
\footnotesize
\begin{tabularx}{\linewidth}{p{1.8cm}Xp{1.5cm}p{1.4cm}}
\toprule
Symbol & Description & Domain & Step \\
\midrule
$H_\mathrm{capacity}$           & Gross time budget per FTE per sprint                                          & $\mathbb{R}_{>0}$     & 1 \\[2pt]
$H_\mathrm{mon,overhead}$       & Monitoring hours consumed per supervised agent                       & $\mathbb{R}_{>0}$     & 1 \\[2pt]
$H_\mathrm{review\_time}$       & Hours consumed per architecture proposal reviewed                 & $\mathbb{R}_{>0}$     & 1 \\[2pt]
$H_\mathrm{esc\_time}$          & Hours consumed per escalation handled                             & $\mathbb{R}_{>0}$     & 1 \\[2pt]
$h_\mathrm{feature}$            & Human hours required to implement one feature                     & $\mathbb{R}_{>0}$     & 1 \\[2pt]
$h_\mathrm{fix}$                & Human hours required to fix one bug                               & $\mathbb{R}_{>0}$     & 1 \\[2pt]
$c_\mathrm{comm}$               & Communication overhead hours per developer per additional peer       & $\mathbb{R}_{\geq 0}$ & 1 \\
\bottomrule
\end{tabularx}
\end{table}

\begin{table}[t]
\caption{Feature-pipeline and throughput-rate parameters.}
\label{tab:params:pipeline}
\footnotesize
\begin{tabularx}{\linewidth}{p{1.8cm}Xp{1.5cm}p{1.4cm}}
\toprule
Symbol & Description & Domain & Step \\
\midrule
$r_\mathrm{arrival}$   & Poisson rate of feature arrivals per sprint                          & $\mathbb{R}_{>0}$  & 3 \\[2pt]
$r_F$                  & Features implemented per agent per sprint                                                & $\mathbb{R}_{>0}$  & 3 \\[2pt]
$r_\mathrm{arch,chg}$  & Architecture change requests resolved per Architecture Agent per sprint                 & $\mathbb{R}_{>0}$  & 6 \\[2pt]
$r_\mathrm{refactor}$  & Debt reduction per Refactoring Agent per sprint        & $\mathbb{R}_{>0}$  & 9 \\[2pt]
$r_\mathrm{test}$      & Tests written per agent per sprint                    & $\mathbb{R}_{>0}$  & 4 \\[2pt]
$r_\mathrm{doc}$       & Documentation items written per agent per sprint    & $\mathbb{R}_{>0}$  & 9 \\[2pt]
$r_\mathrm{fix}$       & Bugs fixed per Bug Fixing Agent per sprint            & $\mathbb{R}_{>0}$  & 7 \\
\bottomrule
\end{tabularx}
\end{table}

\begin{table}[t]
\caption{Bug-injection parameters.}
\label{tab:params:injection}
\footnotesize
\begin{tabularx}{\linewidth}{p{1.8cm}Xp{1.5cm}p{1.4cm}}
\toprule
Symbol & Description & Domain & Step \\
\midrule
$r_\mathrm{bugs}$               & Base Poisson rate of bugs injected per feature by agents                                         & $\mathbb{R}_{>0}$     & 5 \\[2pt]
$r_\mathrm{bugs,human}$         & Poisson rate of bugs injected per feature by humans                      & $\mathbb{R}_{>0}$               & 5 \\[2pt]
$\theta_\mathrm{maj}$           & Share of major bugs among all injected bugs & $(0,1)$               & 5 \\[2pt]
$\theta_\mathrm{sec}$           & Share of security bugs among all injected bugs & $(0,1)$               & 5 \\[2pt]
\bottomrule
\end{tabularx}
\end{table}

\begin{table}[t]
\caption{Bug detection-rate and quality-degradation parameters.}
\label{tab:params:detection}
\footnotesize
\begin{tabularx}{\linewidth}{p{1.8cm}Xp{1.5cm}p{1.4cm}}
\toprule
Symbol & Description & Domain & Step \\
\midrule
$d_\mathrm{CR}$                 & Base binomial detection probability for Code Review                                    & $[0,1]$               & 5 \\[2pt]
$d_\mathrm{SA}$                 & Base binomial detection probability for Security Audit                                 & $[0,1]$               & 5 \\[2pt]
$d_\mathrm{test}$               & Maximum test detection probability at full coverage                     & $[0,1]$               & 5 \\[2pt]
$c_\mathrm{test,ceil}$          & Upper bound on test coverage                                                           & $(0,1]$               & 4 \\[2pt]
$\alpha_\mathrm{size}$          & Detection degradation based on change volume per review/audit agent & $\mathbb{R}_{\geq 0}$ & 5 \\[2pt]
\bottomrule
\end{tabularx}
\end{table}

\begin{table}[t]
\caption{Debt, architecture, escalation, and documentation parameters.}
\label{tab:params:debt}
\footnotesize
\begin{tabularx}{\linewidth}{p{1.8cm}Xp{1.5cm}p{1.4cm}}
\toprule
Symbol & Description & Domain & Step \\
\midrule
$\delta_\mathrm{min}$           & Floor of debt amplifier $\delta_D$ at $D_\mathrm{eff}=1$, i.e., lower bound for throughput, review, and escalation quality & $(0,1)$ & 2 \\[2pt]
$\delta_\mathrm{exp}$           & Shape exponent of the debt amplifier $\delta_D$                          & $\mathbb{R}_{>0}$     & 2 \\[2pt]
$f_\mathrm{arch,chg}$  & Fraction of implemented features that generate an architecture change request & $[0,1]$            & 6 \\[2pt]
$\delta_\mathrm{impl}$          & Additive debt increase per feature implemented        & $\mathbb{R}_{>0}$     & 9 \\[2pt]
$\delta_\mathrm{fix}$           & Additive debt increase per bug repaired               & $\mathbb{R}_{\geq 0}$ & 9 \\[2pt]
$\rho_\mathrm{esc}$             & Escalation probability per completed agent work item                                     & $\mathbb{R}_{\geq 0}$ & 8 \\[2pt]
$\Gamma_\mathrm{doc}$           & Maximum fractional reduction of monitoring overhead at full documentation coverage & $[0,1]$ & 1 \\
\bottomrule
\end{tabularx}
\end{table}

\begin{table}[t]
\caption{Cost parameters. $c_\mathrm{token}$ is an 8-element tuple, one entry per agent role.}
\label{tab:params:costs}
\footnotesize
\begin{tabularx}{\linewidth}{p{1.8cm}Xp{1.5cm}p{1.4cm}}
\toprule
Symbol & Description & Domain & Section \\
\midrule
$c_\mathrm{token}$              & API token cost per deployed agent, indexed by role, per unit of the role's throughput rate $r_k$ & $\mathbb{R}_{\geq 0}^8$ & \ref{sec:model:objectives} \\[2pt]
$c_\mathrm{salary}$             & Annual fully-loaded salary of one FTE                & $\mathbb{R}_{\geq 0}$   & \ref{sec:model:objectives} \\[2pt]
$\omega_\mathrm{min}$           & Expected cost of one escaped minor bug                                       & $\mathbb{R}_{>0}$        & \ref{sec:model:objectives} \\[2pt]
$\omega_\mathrm{maj}$           & Expected cost of one escaped major bug                                       & $\mathbb{R}_{>0}$        & \ref{sec:model:objectives} \\[2pt]
$\omega_\mathrm{sec}$           & Expected cost of one escaped security bug                                    & $\mathbb{R}_{>0}$        & \ref{sec:model:objectives} \\
\bottomrule
\end{tabularx}
\end{table}

% ---- 3.6 ----
\subsection{Sprint transition function}
\label{sec:model:transition}

The sprint transition $\mathcal{T}: S(t-1) \to S(t)$ executes ten steps in
fixed dependency order. 

\subsubsection*{Step 1 --- Human capacity allocation}

The priority-ordered time budget of the developers is
resolved as a five-stage cascade, each stage consuming the time left after the previous one. First, communication overhead between the developers reduces capacity to
\begin{equation}
  H_\mathrm{net} = H_\mathrm{capacity} - c_\mathrm{comm}\lceil N-1 \rceil
  \label{eq:hnet}
\end{equation}
per FTE. Communication is a per-person cost that does not shrink when a person invests only
part of their time in the project, so the overhead is driven by the headcount rather than by
the staffing level: $\lceil N \rceil$ is the smallest headcount that can supply $N$ FTE,
so $\lceil N-1 \rceil$ is the number of peers each of them communicates with. Since $n_\mathrm{sup}(t)$ is the number of active supervised agents, each FTE supervises
$\frac{n_\mathrm{sup}(t)}{N}$ agents, consuming $\frac{n_\mathrm{sup}(t)}{N} \cdot H_\mathrm{mon,eff}(t)$ hours of monitoring effort. The per-agent monitoring effort is reduced by the documentation coverage $\mathrm{cov\_doc}(t-1)$ (from the previous sprint) and the reduction is calibrated by $\Gamma_\mathrm{doc}$
\begin{equation}
  H_\mathrm{mon,eff}(t) = H_\mathrm{mon,overhead}\,
    \bigl(1 - \Gamma_\mathrm{doc}\,\mathrm{cov\_doc}(t-1)\bigr),
  \label{eq:hmoneff}
\end{equation}
leaving post-monitoring capacity
\begin{equation}
  H_\mathrm{after\_mon}(t) = \max\!\Bigl(0,\; H_\mathrm{net} -
    \frac{n_\mathrm{sup}(t)}{N}\,H_\mathrm{mon,eff}(t)\Bigr).
  \label{eq:hafter_mon}
\end{equation}
We assume $\Gamma_\mathrm{doc} \leq 1$, so that documentation reduces, but never inverts, the monitoring effort.

Architecture reviews are served from this remainder and drain the architecture review queue $Q_\mathrm{rev}(t-1)$ at rate
$N/H_\mathrm{review\_time}$:
\begin{align}
  \mathrm{rev\_done}(t) &= \min\!\Bigl(
    \bigl\lfloor H_\mathrm{after\_mon}(t)\cdot N\,/\,H_\mathrm{review\_time}\bigr\rfloor,\;
    Q_\mathrm{rev}(t-1)\Bigr),
  \label{eq:rev_done}
\end{align}
leaving post-review capacity
\begin{equation}
  H_\mathrm{after\_rev}(t) =
    H_\mathrm{after\_mon}(t) - \mathrm{rev\_done}(t)\,H_\mathrm{review\_time}/N.
  \label{eq:hafter_rev}
\end{equation}
Escalations are served analogously:
\begin{equation}
  \mathrm{esc\_done}(t) = \min\!\Bigl(
    \bigl\lfloor H_\mathrm{after\_rev}(t)\cdot N\,/\,H_\mathrm{esc\_time}\bigr\rfloor,\;
    Q_\mathrm{esc}(t-1)\Bigr),
  \label{eq:esc_done}
\end{equation}
leaving post-escalation capacity
\begin{equation}
  H_\mathrm{after\_esc}(t) =
    H_\mathrm{after\_rev}(t) - \mathrm{esc\_done}(t)\,H_\mathrm{esc\_time}/N.
  \label{eq:hafter_esc}
\end{equation}
The remaining capacity $H_\mathrm{after\_esc}(t)$ is used for coding:
fraction $f_\mathrm{fix}$ goes to bug fixing and $(1-f_\mathrm{fix})$ to feature implementation:
\begin{align}
  \mathrm{fix\_human}(t)  &=
    \bigl\lfloor N\,f_\mathrm{fix}\,H_\mathrm{after\_esc}(t)\,/\,h_\mathrm{fix}\bigr\rfloor,
  \label{eq:fixhuman} \\
  \mathrm{impl\_human}(t) &=
    \bigl\lfloor N\,(1-f_\mathrm{fix})\,H_\mathrm{after\_esc}(t)\,/\,h_\mathrm{feature}\bigr\rfloor.
  \label{eq:implhuman}
\end{align}
During feature-freeze sprints (Section~\ref{sec:model:cycle}), no features are implemented, i.e.,
$\mathrm{impl\_human}=0$, and all of $H_\mathrm{after\_esc}(t)$ is redirected to bug fixing:
\begin{equation}
  \mathrm{fix\_human}(t)\big|_\mathrm{freeze} =
    \bigl\lfloor N\,H_\mathrm{after\_esc}(t)\,/\,h_\mathrm{fix}\bigr\rfloor.
  \label{eq:fixhuman_freeze}
\end{equation}
Both activities are debt-independent: $h_\mathrm{feature}$ and $h_\mathrm{fix}$ are fixed
costs regardless of the current debt level $D_\mathrm{tech}(t-1)$. While not fully realistic, this simplifies the model, and we assume that the human effort already includes dealing with possibly existing debt-related complexity. 

$H_\mathrm{after\_mon}(t)$ is driven to zero if monitoring overhead exceeds $H_\mathrm{net}$.
Configurations for which this is unavoidable cannot perform any productive human work regardless of the scenario-parameter draws. We therefore define a feasibility criterion that is checked before running the simulation for the development composition $\mathbf{A}_\mathrm{dev}$, and for the QA composition $\mathbf{A}_\mathrm{qa}$ when $F_\mathrm{freeze} \geq 1$:
\begin{align}
  H_\mathrm{net} = H_\mathrm{capacity} - c_\mathrm{comm}\lceil N-1 \rceil &\geq 0 \\
  \frac{\sum_{k \in \mathcal{K}_\mathrm{sup}} a_{\cdot,k}}{N}\,H_\mathrm{mon,overhead}\,
    (1 - \Gamma_\mathrm{doc}) &< H_\mathrm{net}.
  \label{eq:feas_monitoring}
\end{align}
This check ensures that the developers have sufficient capacity for the communication between
themselves and for the monitoring of agents under the best-case assumption of perfect documentation coverage. The criterion is a necessary condition rather than a sufficient one: since the documentation coverage starts below its ceiling, the effective monitoring overhead early in a simulation run can still temporarily exceed the available capacity. We do not specify similar constraints for architecture review and escalation time, as not handling these only increases the debt penalties we discuss next, meaning that implementation is still possible, though less efficient.

\subsubsection*{Step 2 --- Debt penalties}

Technical debt $D_\mathrm{tech}$, as well as unresolved work in the architecture
and escalation pipelines, contributes to productivity degradation.
Architecture work is considered unresolved as long as it remains anywhere in
the pipeline: as an unprocessed change request ($Q_\mathrm{arch}$), as a design
proposal awaiting human review ($Q_\mathrm{rev}$), or as an approved proposal
awaiting implementation ($Q_\mathrm{arch,impl}$). We normalize this by the team's \emph{maximum throughput} for the
corresponding human activity in one sprint
$H_\mathrm{capacity} \cdot N \;/\; H_{x}$.
A value of~1 means the queue holds as many items as the team could process in
an entire sprint under ideal conditions, i.e., the team would need a full sprint just to drain the queue. Once the queue reaches this or a worse state, the architectural degradation has reached its maximum. Step~2 reads queue values carried over from sprint $t-1$, so all debt quantities use the $t-1$ index. Overall, this gives us the architectural debt as
\begin{equation}
  D_\mathrm{arch}(t-1) = \min\!\Bigl(1,\;
    \frac{Q_\mathrm{arch}(t-1) + Q_\mathrm{rev}(t-1) + Q_\mathrm{arch,impl}(t-1)}
         {H_\mathrm{capacity}\,N \;/\; H_\mathrm{review\_time}}\Bigr).
  \label{eq:D_arch}
\end{equation}
Similarly, a long escalation queue indicates that the team is overloaded, because unresolved
escalations mean that agents require frequent human intervention, which
degrades coordination quality.  Normalizing by the maximum escalation throughput
$H_\mathrm{capacity}\,N / H_\mathrm{esc\_time}$ gives
\begin{equation}
  D_\mathrm{esc}(t-1) = \min\!\Bigl(1,\;
    \frac{Q_\mathrm{esc}(t-1)}
         {H_\mathrm{capacity}\,N \;/\; H_\mathrm{esc\_time}}\Bigr).
  \label{eq:D_esc}
\end{equation}
The combined effective debt is
\begin{equation}
  D_\mathrm{eff}(t-1) = \min\!\bigl(1,\; D_\mathrm{tech}(t-1) + D_\mathrm{arch}(t-1) + D_\mathrm{esc}(t-1)\bigr).
  \label{eq:D_eff}
\end{equation}
The \emph{debt amplifier}
\begin{equation}
  \delta_D = 1 - (1 - \delta_\mathrm{min})\,D_\mathrm{eff}(t-1)^{\delta_\mathrm{exp}}
  \label{eq:delta_D}
\end{equation}
is a single function of the combined debt, parameterized by a floor
$\delta_\mathrm{min} \in (0,1)$ that describes the highest negative impact the debt can have and a shape exponent $\delta_\mathrm{exp} > 0$ that controls the rate of decay.
It equals~1 at zero combined debt and decays to $\delta_\mathrm{min}$ at
$D_\mathrm{eff} = 1$. This debt multiplier will be used in subsequent steps to reduce feature implementation speed, increase bug injection, reduce code review quality, slow bug fixing, and increase the rate at which agents require escalations.

\subsubsection*{Step 3 --- Feature arrival and implementation}
In every sprint, new features arrive in the backlog according to
\begin{equation}
  \mathrm{arrivals}(t) \sim \mathrm{Poisson}(r_\mathrm{arrival}),
  \label{eq:arrivals}
\end{equation}
growing the backlog at the beginning of the step. Because arrivals are added before implementation runs, newly arrived features are immediately available for implementation in the same sprint. The backlog is then drained by human and agent implementation.

Available agent capacity is first consumed by approved architecture proposals
$Q_\mathrm{arch,impl}(t-1)$ pending implementation, with remaining capacity applied
to the feature backlog. Writing
$\mathrm{demand}(t) = F_\mathrm{backlog}(t-1) + \mathrm{arrivals}(t)$ for the features
available for implementation, the human developers are served first and the agents
implement what remains. The rationale is that the developers' salaries are paid regardless, so their capacity should be used first.
\begin{align}
  \mathrm{arch\_impl}(t) &= \min\!\Bigl(Q_\mathrm{arch,impl}(t-1),\;
    \bigl\lfloor r_F\,\delta_D\,a_{\cdot,1}/5\bigr\rfloor\Bigr),
  \label{eq:arch_impl} \\
  \mathrm{impl\_human}'(t) &= \min\!\bigl(\mathrm{impl\_human}(t),\;\mathrm{demand}(t)\bigr),
  \label{eq:impl_human_capped} \\
  \mathrm{impl\_agents}(t) &= \min\!\Bigl(\mathrm{demand}(t) - \mathrm{impl\_human}'(t),\;
    \nonumber \\
  &\qquad\bigl\lfloor r_F\,\delta_D\,a_{\cdot,1}\bigr\rfloor - 5\,\mathrm{arch\_impl}(t)\Bigr).
  \label{eq:impl_agents}
\end{align}
Implementing approved architecture proposals ($\mathrm{arch\_impl}$) consumes agent
capacity but does not increment $F_\mathrm{completed}$, since it represents internal
architecture work rather than user-visible feature delivery. Each such implementation
consumes five features' worth of agent capacity, because it requires the same
code-base-wide analysis that makes an architecture proposal five times as expensive to
produce (Section~\ref{sec:scenario:agentcosts}). The feature implementation is linearly reduced by $\delta_D$ to reflect the impact of debt on development speed.
Total user-feature implementation per sprint is capped at the available backlog to
prevent over-counting features:
\begin{equation}
  \mathrm{impl}(t) = \mathrm{impl\_agents}(t) + \mathrm{impl\_human}'(t)
    \;\leq\; \mathrm{demand}(t),
  \label{eq:impl}
\end{equation}
and the state variables are updated as:
\begin{align}
  F_\mathrm{backlog}(t)   &= F_\mathrm{backlog}(t-1) + \mathrm{arrivals}(t) - \mathrm{impl}(t),
    \label{eq:backlog_update} \\
  F_\mathrm{completed}(t) &= F_\mathrm{completed}(t-1) + \mathrm{impl}(t),
    \label{eq:completed_update} \\
  F_\mathrm{cycle}(t) &= F_\mathrm{cycle}(t-1) + \mathrm{impl}(t).
    \label{eq:cycle_update}
\end{align}

\subsubsection*{Step 4 --- Test suite update}
The Test Development Agent grows the
cumulative test count before bug detection runs, up to the coverage ceiling of
Eq.~\eqref{eq:coverage}, beyond which further tests no longer raise
$\mathrm{cov\_test}$:
\begin{equation}
  N_\mathrm{test}(t) = \min\!\bigl(N_\mathrm{test}(t-1) + r_\mathrm{test}\,a_{\cdot,7},\;
    c_\mathrm{test,ceil}\,F_\mathrm{completed}(t)\bigr).
  \label{eq:ntest_update}
\end{equation}
Because this step executes before bug injection and detection (Step~5), the new
tests written this sprint are immediately available to raise the test coverage
$\mathrm{cov\_test}$. Thus, investments in testing take effect in the same sprint, which directly incentivizes the allocation of Test Development Agents.

\subsubsection*{Step 5 --- Bug injection and detection}
Agents inject bugs at an effective per-feature bug introduction rate
\begin{equation}
  r_\mathrm{bugs,eff}(t) = r_\mathrm{bugs}\,(2 - \delta_D),
  \label{eq:r_bugs_eff}
\end{equation}
where the factor $(2 - \delta_D) \in [1,\, 2 - \delta_\mathrm{min}]$ increases the bug injection as debt accumulates. How many bugs exactly are introduced is randomized using a Poisson distribution as
\begin{equation}
  B_\mathrm{agents}(t) \sim
    \mathrm{Poisson}\!\bigl(\mathrm{impl\_agents}(t)\cdot r_\mathrm{bugs,eff}(t)\bigr).
  \label{eq:Binj-agent}
\end{equation}
The human-implemented features inject bugs at a fixed per-feature rate, i.e., independent of the debt as
\begin{equation}
  B_\mathrm{human}(t)  \sim
    \mathrm{Poisson}\!\bigl(\mathrm{impl\_human}'(t)\cdot r_\mathrm{bugs,human}\bigr),
  \label{eq:Binj-human}
\end{equation}
which gives us
\begin{equation}
  B_\mathrm{inj}(t) = B_\mathrm{agents}(t) + B_\mathrm{human}(t)
  \label{eq:Binj-total}
\end{equation}
as the overall number of injected bugs within a sprint. 

Each injected bug is assigned to exactly one of three mutually exclusive types via
a sequential draw: security bugs are extracted first, then the remainder is split
into major and minor:
\begin{align}
  B_\mathrm{sec}(t) &\sim \mathrm{Binomial}\!\bigl(B_\mathrm{inj}(t),\;\theta_\mathrm{sec}\bigr),
    \nonumber \\
  B_\mathrm{maj}(t) &\sim \mathrm{Binomial}\!\Bigl(B_\mathrm{inj}(t)-B_\mathrm{sec}(t),\;
    \tfrac{\theta_\mathrm{maj}}{\theta_\mathrm{min}+\theta_\mathrm{maj}}\Bigr),
    \nonumber \\
  B_\mathrm{min}(t) &= B_\mathrm{inj}(t) - B_\mathrm{sec}(t) - B_\mathrm{maj}(t),
  \label{eq:type_split}
\end{align}
where $\theta_\mathrm{min} = 1 - \theta_\mathrm{maj} - \theta_\mathrm{sec}$.

Detection proceeds in three sequential stages that model a workflow where first an existing test suite is run against code
changes (simulating a CI pipeline), then code review is conducted, and finally a security audit is performed by a specialist agent. 
Which bugs are inspected depends on the phase of the release cycle. During development sprints, the detection inspects the features developed in the current sprint and can hence only detect bugs in those features. The test suite is run first, and the change volume under inspection consists of the features implemented in the sprint. During feature-freeze sprints, the whole code of the current release cycle is re-inspected to detect bugs that are still hidden. This gives us the inspected bugs and the inspected volume as
\begin{align}
  P_x(t) &=
  \begin{cases}
    B_x(t) & \text{development sprints},\\[2pt]
    B_\mathrm{hid,x}(t-1) & \text{feature-freeze sprints},
  \end{cases}
  \nonumber \\[4pt]
  v(t) &=
  \begin{cases}
    \mathrm{impl}(t) & \text{development sprints},\\[2pt]
    F_\mathrm{cycle}(t) & \text{feature-freeze sprints},
  \end{cases}
  \label{eq:inspected}
\end{align}
for $x \in \{\mathrm{min,maj,sec}\}$.

The test suite detects bugs at rate
\begin{equation}
  p_\mathrm{test}(t) =
  \begin{cases}
    \min\!\bigl(1,\; d_\mathrm{test}\,\mathrm{cov\_test}(t)\bigr) & \text{development sprints},\\[2pt]
    0 & \text{feature-freeze sprints},
  \end{cases}
  \label{eq:p_test}
\end{equation}
where $\mathrm{cov\_test}(t)$ is given by Eq.~\eqref{eq:coverage}. The test suite is not re-run during feature freezes, since it already ran against every change when it was implemented.

Within our model, we assume that both agents invest effort that is reasonable to review a single feature in depth. If more features need to be reviewed, the per-feature review effectiveness is reduced. We model this through a degradation function $\Phi$ of the number of features relative to the number of agents. We use a logarithmic transformation to model this as diminishing returns:
\begin{equation}
  \Phi(v, a) =
  \begin{cases}
    \max\!\Bigl(0,\; 1 - \alpha_\mathrm{size}\log\!\bigl(\tfrac{v}{a} + 1\bigr)\Bigr) & a \geq 1,\\[4pt]
    0 & a = 0.
  \end{cases}
  \label{eq:volume_factor}
\end{equation}
The code review then detects bugs that survive testing at effective rate
\begin{equation}
  d_\mathrm{CR,eff}(t) = d_\mathrm{CR}\,\delta_D\,\Phi\bigl(v(t),\, a_{\cdot,5}\bigr),
  \label{eq:dcr_eff}
\end{equation}
and the Security Audit Agent detects the security bugs that survive both earlier stages at rate
\begin{equation}
  d_\mathrm{SA,eff}(t) = d_\mathrm{SA}\,\delta_D\,\Phi\bigl(v(t),\, a_{\cdot,6}\bigr).
  \label{eq:dsa_eff}
\end{equation}

Let $T_x(t)$, $\mathrm{CR}_x(t)$ denote the counts detected by tests and code review for type
$x$, and $A_\mathrm{sec}(t)$ the count detected by the security audit. These are given by binomial draws with the corresponding detection probabilities:
\begin{align}
  T_x(t)        &\sim \mathrm{Binomial}\!\bigl(P_x(t),\; p_\mathrm{test}(t)\bigr),
    \quad x \in \{\mathrm{min,maj,sec}\}, \nonumber \\
  \mathrm{CR}_x(t)  &\sim \mathrm{Binomial}\!\bigl(P_x(t)-T_x(t),\; d_\mathrm{CR,eff}(t)\bigr),
    \quad x \in \{\mathrm{min,maj,sec}\}, \nonumber \\
  A_\mathrm{sec}(t) &\sim \mathrm{Binomial}\!\bigl(P_\mathrm{sec}(t)-T_\mathrm{sec}(t)-\mathrm{CR}_\mathrm{sec}(t),\; d_\mathrm{SA,eff}(t)\bigr).
    \nonumber
\end{align}
The known and hidden queues are incremented accordingly:
\begin{align}
  B_\mathrm{kno,x}(t)   &= B_\mathrm{kno,x}(t-1) + T_x(t) + \mathrm{CR}_x(t),
    \quad x \in \{\mathrm{min,maj}\}, \label{eq:bkno_update} \\
  B_\mathrm{kno,sec}(t) &= B_\mathrm{kno,sec}(t-1) + T_\mathrm{sec}(t) + \mathrm{CR}_\mathrm{sec}(t) + A_\mathrm{sec}(t),
    \nonumber \\
  B_\mathrm{hid,x}(t)   &= B_\mathrm{hid,x}(t-1) + B_x(t) - T_x(t) - \mathrm{CR}_x(t),
    \quad x \in \{\mathrm{min,maj}\}, \nonumber \\
  B_\mathrm{hid,sec}(t) &= B_\mathrm{hid,sec}(t-1) + B_\mathrm{sec}(t) - T_\mathrm{sec}(t) - \mathrm{CR}_\mathrm{sec}(t) - A_\mathrm{sec}(t).
    \nonumber
\end{align}
Note that $B_\mathrm{kno,x}(t)$ is an intermediate value at this step, which will be decreased by the bug fixing conducted in Step~7. 

\subsubsection*{Step 6 --- Architecture queues}
This step propagates architecture work through three queues $Q_\mathrm{arch}$, $Q_\mathrm{rev}$, and $Q_\mathrm{arch,impl}$ in causal order. First, we determine how many new architecture change requests are generated by the features implemented this sprint.
Architecture change requests are generated at rate $f_\mathrm{arch,chg}$ per
feature implemented this sprint,
giving Poisson-distributed arrivals that fill the change-request queue:
\begin{align}
  \mathrm{arch\_chg\_new}(t) &\sim \mathrm{Poisson}\!\bigl(f_\mathrm{arch,chg}\cdot\mathrm{impl}(t)\bigr)
  \label{eq:arch_chg_new} \\
  Q_\mathrm{arch}(t) &= Q_\mathrm{arch}(t-1) + \mathrm{arch\_chg\_new}(t).
  \label{eq:Q_arch_new}
\end{align}

Then, the Architecture Agents drain $Q_\mathrm{arch}$ at rate $r_\mathrm{arch,chg}$ per agent.
Each resolved change request produces one design proposal forwarded to the human
review queue $Q_\mathrm{rev}$, which is at the same time reduced by the human review capacity:
\begin{align}
  \mathrm{arch\_chg\_done}(t) &= \min\!\bigl(Q_\mathrm{arch}(t),\;
    \lfloor r_\mathrm{arch,chg}\,a_{\cdot,2}\rfloor\bigr), \label{eq:arch_chg_done} \\
  Q_\mathrm{arch}(t) &= Q_\mathrm{arch}(t) - \mathrm{arch\_chg\_done}(t),
    \label{eq:Q_arch} \\
  Q_\mathrm{rev}(t) &= \max\!\bigl(0,\;
    Q_\mathrm{rev}(t-1) + \mathrm{arch\_chg\_done}(t) - \mathrm{rev\_done}(t)\bigr).
    \label{eq:Q_rev}
\end{align}

Finally, the $\mathrm{rev\_done}(t)$ proposals reviewed and approved this sprint are placed
into the \emph{architecture implementation queue} $Q_\mathrm{arch,impl}$, enabling their implementation in the next sprint by the Feature Implementation Agents in Step~3:
\begin{equation}
  Q_\mathrm{arch,impl}(t) = \max\!\bigl(0,\;
    Q_\mathrm{arch,impl}(t-1) + \mathrm{rev\_done}(t) - \mathrm{arch\_impl}(t)\bigr).
  \label{eq:Q_arch_impl}
\end{equation}

\subsubsection*{Step 7 --- Bug fixing}
Bug fixing proceeds in two phases using the same six priority tiers
($B_\mathrm{rel,sec} \to B_\mathrm{kno,sec} \to B_\mathrm{rel,maj} \to
B_\mathrm{kno,maj} \to B_\mathrm{rel,min} \to B_\mathrm{kno,min}$):

\emph{Phase 1: Human fixes (debt-independent).} Human developers apply their
bug-fixing budget $\mathrm{fix\_human}(t)$ to the priority tiers in order. Each bug fixed
consumes one unit of the integer budget, so:
\begin{equation}
  n^H_i(t) = \min\!\Bigl(\text{queue}_i(t),\; B^H_i(t)\Bigr),
  \qquad B^H_{i+1}(t) = B^H_i(t) - n^H_i(t),
  \label{eq:human_fixes}
\end{equation}
where $B^H_1(t) = \mathrm{fix\_human}(t)$.

\emph{Phase 2: Agent fixes (debt-slowed).} The Bug Fixing Agents then drain the
remaining queues, traversing the same tiers in the same order with a shared budget
\begin{equation}
  B^A_1(t) = \bigl\lfloor a_{\cdot,3}\,\delta_D\,r_\mathrm{fix}\bigr\rfloor,
  \label{eq:fix_budget}
\end{equation}
i.e., the per-agent fix rate $r_\mathrm{fix}$ scaled by the number of deployed Bug Fixing
Agents and by the debt amplifier $\delta_D$. As for the humans, each bug fixed consumes one
unit of this budget, irrespective of its severity:
\begin{align}
  n^A_i(t) &= \min\!\Bigl(\text{queue}_i(t) - n^H_i(t),\; B^A_i(t)\Bigr),
  \label{eq:fixes} \\
  B^A_{i+1}(t) &= B^A_i(t) - n^A_i(t). \nonumber
\end{align}
The two phases therefore differ only in how their budget is derived: the human budget follows
from the available human capacity and is debt-independent, whereas the agent budget scales with
the number of deployed agents and shrinks as technical debt accumulates.
Since human fixes take priority, a well-staffed team can
keep the known-bug queue short even when $\delta_D$ is low.

Letting $n_i(t) = n^H_i(t) + n^A_i(t)$ denote the total bugs fixed from tier $i$
across both phases, the release and known-bug queues are decremented:
\begin{align}
  B_\mathrm{rel,sec}(t) &= B_\mathrm{rel,sec}(t-1) - n_1(t), \nonumber \\
  B_\mathrm{kno,sec}(t) &= B_\mathrm{kno,sec}(t)   - n_2(t), \nonumber \\
  B_\mathrm{rel,maj}(t) &= B_\mathrm{rel,maj}(t-1) - n_3(t), \nonumber \\
  B_\mathrm{kno,maj}(t) &= B_\mathrm{kno,maj}(t)   - n_4(t), \nonumber \\
  B_\mathrm{rel,min}(t) &= B_\mathrm{rel,min}(t-1) - n_5(t), \nonumber \\
  B_\mathrm{kno,min}(t) &= B_\mathrm{kno,min}(t)   - n_6(t).
  \label{eq:bfix_update}
\end{align}
The $B_\mathrm{rel,x}$ rows use index $t-1$ on the right because the release
queues are not written by any earlier step in this sprint.  The $B_\mathrm{kno,x}$
rows use index $t$ as a reminder that their right-hand side values
were already updated by Step~5.

\subsubsection*{Step 8 --- Escalation queue}
Agents raise human escalations per completed work item at rate $\rho_\mathrm{esc}$.
The work the supervised implementation agents completed this sprint is
\begin{equation}
  W_\mathrm{agent}(t) = \mathrm{impl\_agents}(t) + \mathrm{arch\_chg\_done}(t)
    + \mathrm{fixes\_agent}(t),
  \label{eq:w_agent}
\end{equation}
where $\mathrm{fixes\_agent}(t) = \sum_{i=1}^{6} n^A_i(t)$ is the agent share of the
repairs of Step~7. Human work is not included, since developers do not escalate to
themselves. The two review services are also excluded, because they report their findings as bugs instead of
escalating. The same holds for the test and documentation agents, whose artifacts are reviewed
together with the code they belong to. New arrivals each sprint are then
$\mathrm{esc\_new}(t) \sim \mathrm{Poisson}\bigl(\rho_\mathrm{esc}\,(2-\delta_D)\,W_\mathrm{agent}(t)\bigr)$,
where the factor $(2 - \delta_D)$ increases the escalation rate as debt accumulates, as
it does for bug injection in Eq.~\eqref{eq:r_bugs_eff}. The queue is then updated taking
the capacity for human escalation resolution $\mathrm{esc\_done}(t)$ into account:
\begin{equation}
  Q_\mathrm{esc}(t) = \max\!\bigl(0,\;
    Q_\mathrm{esc}(t-1) + \mathrm{esc\_new}(t) - \mathrm{esc\_done}(t)\bigr).
  \label{eq:qesc}
\end{equation}

\subsubsection*{Step 9 --- General maintenance}
After the feature implementation, testing, and bug fixing steps, the general maintenance
agents perform their work. They do not provide direct utilitarian value in the form of features delivered or bugs fixed, but instead provide indirect value by slowing the accumulation of technical debt and by improving the documentation, which reduces the cognitive load of the human developers when supervising agents.

The first such activity is performed by the Refactoring Agent, which reduces technical debt. 
Three activities drive the technical debt $D_\mathrm{tech}(t)$: feature implementation and
bug fixing each \emph{increase} debt, while refactoring conducted by the Refactoring Agent 
\emph{decreases} it.
\begin{equation}
\begin{split}
  D_\mathrm{tech}(t) = \operatorname{clip}\bigl(D_\mathrm{tech}(t-1)
    &+ \delta_\mathrm{impl}\,\mathrm{impl}(t) \\
    &+ \delta_\mathrm{fix}\,\mathrm{fixes\_total}(t) \\
    &- r_\mathrm{refactor}\,a_{\cdot,4},\; 0,\, 1\bigr),
  \label{eq:debt_update}
\end{split}
\end{equation}
where $\mathrm{fixes\_total}(t) = \sum_{i=1}^{6} n_i(t)$ is the total number of bugs fixed across all priority tiers in Step~7.

The second activity is documentation written by the Developer Documentation Agent, which grows the cumulative
document count, up to one document per completed feature, i.e., the point at which
Eq.~\eqref{eq:doc_cov} is capped:
\begin{equation}
  N_\mathrm{doc}(t) = \min\!\bigl(N_\mathrm{doc}(t-1) + r_\mathrm{doc}\,a_{\cdot,8},\;
    F_\mathrm{completed}(t)\bigr).
  \label{eq:ndoc_update}
\end{equation}
This increases the documentation coverage $\mathrm{cov\_doc}$ in the next
sprint (see Eq.~\eqref{eq:doc_cov}).

\subsubsection*{Step 10 --- Release event}
If $t \bmod R = 0$, a release fires: all bugs remaining in the hidden and known queues
escape to production: bugs that were never detected as well as bugs that were detected but not fixed before the release deadline.
These bugs are added to the cumulative escaped counter
and to the in-production repair queue. Afterwards, the hidden and known queues are reset to zero, and a new release cycle starts:
\begin{align}
  B_\mathrm{esc,x}(t)  &= B_\mathrm{esc,x}(t-1) + B_\mathrm{hid,x}(t) + B_\mathrm{kno,x}(t),
    \quad x \in \{\mathrm{min,maj,sec}\}, \nonumber \\
  B_\mathrm{rel,x}(t)  &= B_\mathrm{rel,x}(t)   + B_\mathrm{hid,x}(t) + B_\mathrm{kno,x}(t),
    \quad x \in \{\mathrm{min,maj,sec}\}, \nonumber \\
  B_\mathrm{hid,x}(t)  &= 0,
    \quad x \in \{\mathrm{min,maj,sec}\}, \nonumber \\
  B_\mathrm{kno,x}(t)  &= 0,
    \quad x \in \{\mathrm{min,maj,sec}\}, \nonumber \\
  F_\mathrm{cycle}(t) &= 0.
  \label{eq:release_update}
\end{align}
The $B_\mathrm{rel,x}$ rows use index $t$ on the right because the
release queue was already updated by Step~7 (bug fixing) within this sprint.
Similarly, $B_\mathrm{kno,x}(t)$ on the right-hand side is the post-fixing value from
Steps~5 and~7. When $t \bmod R \neq 0$, no release fires and all updates are skipped.

% ---- 3.7 ----
\subsection{Objective functions}
\label{sec:model:objectives}

The model evaluates a configuration through four quantities that are computed from the terminal state $S(T)$ and from the per-sprint record of the work the agents completed, and therefore aggregate performance across the entire development cycle: the \emph{value} delivered by implementing features, and the three cost drivers of the development process, namely the cost of the bugs that escape to production, the cost of the human developers, and the cost of the agents. 

\subsubsection{Value}

The value is defined under the assumption that more features developed mean that the software product is of higher value to its end users. Therefore, we approximate the value produced by the development process by the average throughput of features delivered per sprint, computed as

\begin{equation}
  \mathit{value} = \frac{F_\mathrm{completed}(T)}{T}.
  \label{eq:obj_V}
\end{equation}

\subsubsection{Quality costs}

The quality of software is a second main criterion for end users: if the software is buggy, it will not be useful regardless of the number of features it has. Bugs that escape to production cause costs, e.g., due to outages, security breaches, or at least the effort required to fix them after the release. Therefore, we express quality as the cost that the escaped bugs cause once they affect real users, which is computed as a weighted sum of the escaped bugs of each type:
\begin{equation}
  C_\mathrm{quality} = \omega_\mathrm{min}\,B_\mathrm{esc,min}(T)
          + \omega_\mathrm{maj}\,B_\mathrm{esc,maj}(T)
          + \omega_\mathrm{sec}\,B_\mathrm{esc,sec}(T),
  \label{eq:obj_Q}
\end{equation}
where $\omega_\mathrm{min} \leq \omega_\mathrm{maj} \leq \omega_\mathrm{sec}$
are the expected costs of a single escaped minor, major, and security bug. The counts include all bugs that reached production, i.e., bugs that were never detected as well as bugs that were detected but not fixed
before their release deadline.

\subsubsection{Human personnel costs}

The human personnel cost is the salary of the human developers. The parameter $c_\mathrm{salary}$ represents the annual fully-loaded salary of a human developer, defined for a period of 52 sprints (one sprint per week). The total human cost is computed by multiplying the number of developers $N$ by their salary, pro-rated over the $T$-sprint horizon:
\begin{equation}
  C_\mathrm{human} = N\,c_\mathrm{salary} \cdot \frac{T}{52}.
  \label{eq:cost_human}
\end{equation}

\subsubsection{Agent costs}

Each agent of role $k$ incurs a token cost $c_{\mathrm{token},k}$ per unit of
work produced $r_k$, e.g., per feature added, bug fixed, or test written. Thus, the costs are
computed with respect to the actual throughput and not the theoretical maximum rate. This accounts for under-utilization of agents, e.g., when the feature backlog is empty or when the number of deployed agents exceeds the throughput that the human developers can supervise.
We reuse the rates already introduced for the sprint
transition function where feasible, i.e., where the effort directly depends on the number of features implemented or bugs fixed (Table~\ref{tab:params:pipeline}), namely $r_1 = r_F$,
$r_2 = r_\mathrm{arch,chg}$, $r_3 = r_\mathrm{fix}$,
$r_7 = r_\mathrm{test}$, and $r_8 = r_\mathrm{doc}$. 

Theoretically, we could use the same approach for the other three roles, which would require them to invest a fixed amount of effort per feature implemented, bug fixed, or architecture change request resolved. However, this would be restrictive and not allow the model to explore different quality, debt, and cost trade-offs. Thus, the costs of these agents are flat per-sprint and modeled based on reviewing or addressing the debt of a single feature as the unit. The number of agents deployed then scales costs linearly and the benefit grows with the number of agents, but with diminishing returns once there are many agents per feature added. For these three roles we therefore set
$r_4 = r_5 = r_6 = 1$. 

The five demand-driven roles are therefore charged for the work they actually performed,
i.e., an agent that receives no work causes no costs. In contrast, the three flat-rate roles are
charged per sprint, even when additional agents no longer improve the outcome. Let $u_k(t)$ denote the work role $k$ completed in sprint
$t$, measured in the unit of that role's rate $r_k$ and read off the sprint transition as
summarized in Table~\ref{tab:work_units}. Implementing an approved architecture proposal is
billed as five feature implementations, since it requires the same code-base-wide analysis
that makes an architecture proposal five times as expensive to produce (Section~\ref{sec:scenario:agentcosts}). This gives us the
agent costs
\begin{equation}
  C_\mathrm{agents} = \sum_{k=1}^{8} c_{\mathrm{token},k} \sum_{t=1}^{T} u_k(t).
  \label{eq:cost_agents}
\end{equation}

\begin{table}[t]
\caption{Work unit $u_k(t)$ billed per role and sprint.}
\label{tab:work_units}
\footnotesize
\begin{tabularx}{\linewidth}{cp{2.6cm}X}
\toprule
$k$ & Role & $u_k(t)$ \\
\midrule
1 & Feature Impl.\ & $\mathrm{impl\_agents}(t) + 5\,\mathrm{arch\_impl}(t)$ \\[2pt]
2 & Architecture & $\mathrm{arch\_chg\_done}(t)$ \\[2pt]
3 & Bug Fixing & $\mathrm{fixes\_agent}(t)$ \\[2pt]
4 & Refactoring & $a_{\cdot,4}$ \\[2pt]
5 & Code Review & $a_{\cdot,5}$ \\[2pt]
6 & Security Audit & $a_{\cdot,6}$ \\[2pt]
7 & Test Development & $N_\mathrm{test}(t) - N_\mathrm{test}(t-1)$ \\[2pt]
8 & Dev.\ Documentation & $N_\mathrm{doc}(t) - N_\mathrm{doc}(t-1)$ \\
\bottomrule
\end{tabularx}
\end{table}

% ---- 3.8 ----
\subsection{Monte Carlo protocol}
\label{sec:model:mc}

A feasible configuration $\mathbf{c}$ (cf.\ Section~\ref{sec:model:sosp}) is evaluated by running $K'$
independent Monte Carlo replications. At $t = 0$ all components of $S(0)$ must be initialized. We use fixed values to define the initial state of the system, e.g., zero features in the backlog and zero bugs in all queues.
These values define the state of the project when the first agentic development sprint
starts. They are fixed experiment constants that are
specified as part of the experiment design (see Section~\ref{sec:setup}). The
sprint transition $\mathcal{T}$ is then applied $T$ times, yielding terminal state
$S(T)$. The outcome of each simulation run is randomized by the actions within the sprint transition, which are driven by the stochastic processes, e.g., the Poisson feature arrivals and the binomial bug detection draws.
The value $\mathit{value}_q$ and the three cost drivers $C_{\mathrm{quality},q}$, $C_\mathrm{human}$, and $C_{\mathrm{agents},q}$ for each replication $q=1,\ldots,K'$ are then computed from $S(T)$ and from the per-sprint work record. We combine the cost drivers into the per-replication cost per feature
\begin{equation}
  C_q = \frac{C_{\mathrm{quality},q} + C_\mathrm{human} + C_{\mathrm{agents},q}}
             {F_{\mathrm{completed},q}(T)}
  \label{eq:cost_rep}
\end{equation}
\emph{within} each replication, i.e., the costs of each replication are based on the bugs that escaped, the agent work that was performed, and the features that were delivered in that replication. Note that $C_\mathrm{human}$ carries no replication index, because it depends only on the staffing level $N$ and is therefore deterministic. We use the mean of these values as the estimator for the expected costs $C$. The uncertainty is
quantified with 95\% percentile bootstrap confidence intervals (CIs).

% ---- 3.9 ---- Search Problem
\subsection{Search problem}
\label{sec:model:sosp}

We first define the feasible set $\mathcal{X}$ as the set of all configurations
\[
  \mathbf{c} = (N,\,\mathbf{A}_\mathrm{dev},\,\mathbf{A}_\mathrm{qa},\,R,\,F_\mathrm{freeze},\,f_\mathrm{fix})
\]
satisfying the structural constraints of Section~\ref{sec:model:team} and the
human-capacity feasibility condition of Eq.~\eqref{eq:feas_monitoring}.

A development team has to deliver the work the project demands as cheaply as possible. We model this as a throughput constraint together with a cost objective. The throughput constraint states that the team must keep up with the incoming work: features arrive at a rate of $r_\mathrm{arrival}$ per sprint and cannot be implemented faster than they arrive, so $r_\mathrm{arrival}$ is the expected maximum attainable value. To account for uncertainty of the Poisson arrival of features, we allow a tolerance $\epsilon$ per sprint:
\begin{equation}
  \mathit{value} \;\geq\; (1-\epsilon)\,r_\mathrm{arrival}.
  \label{eq:throughput_constraint}
\end{equation}

Using the feasible configurations that satisfy the throughput constraint, we define our search problem as the minimization of the expected cost per feature delivered (Eq.~\eqref{eq:cost_rep}), i.e., 
\begin{equation}
  \min_{\mathbf{c}\,\in\,\mathcal{X}: \mathit{value} \,\geq\, (1-\epsilon)\,r_\mathrm{arrival}}\; C.
  \label{eq:objective}
\end{equation}

% ---- 3.10 ----
\subsection{Selection under Monte Carlo noise}
\label{sec:model:selection}

Every configuration the search evaluates is scored by a mean over $K'$ Monte Carlo replications. The objective considers only the mean, not the uncertainty of the estimates. We therefore do not report the minimizer alone. Let $\mathbf{c}^\star$ be the configuration with the lowest mean costs that optimizes our constrained search problem with confidence interval $[C^-_{\mathbf{c}^\star},\,C^+_{\mathbf{c}^\star}]$. A second configuration $\mathbf{c}$ that satisfies the throughput constraint and whose confidence interval $[C^-_{\mathbf{c}},\,C^+_{\mathbf{c}}]$ overlaps that of $\mathbf{c}^\star$ is not statistically significantly different at the 95\% confidence level. If no evaluated configuration satisfies the throughput constraint, no such $\mathbf{c}^\star$ exists and $\mathcal{I}^\star$ is empty. In that case we report the configuration with the highest value and mark the result as inadmissible. Otherwise, we define the \emph{indistinguishable set} as
\begin{equation}
  \mathcal{I}^\star = \bigl\{\mathbf{c} \;:\; \overline{\mathit{value}}_{\mathbf{c}} \geq (1-\epsilon)\,r_\mathrm{arrival}
    \;\wedge\; [C^-_{\mathbf{c}},\,C^+_{\mathbf{c}}] \cap [C^-_{\mathbf{c}^\star},\,C^+_{\mathbf{c}^\star}] \neq \emptyset \bigr\}.
  \label{eq:topset}
\end{equation}

Two configurations in $\mathcal{I}^\star$ are statistically indistinguishable by construction, and it can happen that one of them uses no more agents than the other in any role. Reporting the larger of the two would be misleading, since the additional agents do not lead to a measurable improvement. We therefore add a parsimony tie-break within $\mathcal{I}^\star$: we say that $\mathbf{c}$ \emph{parsimony-dominates} $\mathbf{c}'$ iff both lie in $\mathcal{I}^\star$, the two share the same $(N, R, F_\mathrm{freeze}, f_\mathrm{fix})$, $a_{\mathrm{dev},k} \leq a'_{\mathrm{dev},k}$ and $a_{\mathrm{qa},k} \leq a'_{\mathrm{qa},k}$ hold for all $k$ with at least one strict inequality, and the mean cost of $\mathbf{c}$ is strictly lower than that of $\mathbf{c}'$, i.e., $C_\mathbf{c} < C_{\mathbf{c}'}$. Parsimony-dominated configurations are removed from $\mathcal{I}^\star$ before it is reported. The cost condition means that we only drop a larger team in favor of a smaller one that also measured better. In particular, $\mathbf{c}^\star$ has the lowest mean cost within $\mathcal{I}^\star$ by definition and therefore cannot be parsimony-dominated, i.e., $\mathbf{c}^\star$ is always retained.

Both $\mathbf{c}^\star$ and $\mathcal{I}^\star$ are preliminary at this point, because the width of the confidence intervals that Eq.~\eqref{eq:topset} compares is governed by the number of replications $K'$ that the search uses. We re-simulate every configuration in the preliminary $\mathcal{I}^\star$ with $K > K'$ replications to get better estimates of $\mathbf{c}^\star$ and reduce the confidence intervals. Since the tie-break has already removed the parsimony-dominated configurations, only the parsimonious members have to be re-simulated, which reduces the effort of this step. We then apply the selection a second time and expect that $\mathcal{I}^\star$ shrinks, since there should be fewer overlapping confidence intervals. While it is unlikely that the throughput changes, the constraint is checked as well and violating configurations are discarded. Similarly, we also apply the parsimony tie-break a second time, but do not expect major changes.

% ---- 3.11 ----
\subsection{Search strategy}
\label{sec:model:search}

Exhaustive search of the configuration space for the optimal process and team configuration is impossible, because the counts of the two review services are unbounded (Section~\ref{sec:model:team}). Even under an arbitrary cap of 99 agents per service, the four review service counts $a_{\mathrm{dev},5}$, $a_{\mathrm{dev},6}$, $a_{\mathrm{qa},5}$, and $a_{\mathrm{qa},6}$ alone already give $100^4 = 10^8$ combinations. The possible values of the human staffing level, the release cadence, the feature-freeze length, and the supervised parts of both compositions multiply this number further.

We therefore search Eq.~\eqref{eq:objective} heuristically with an \emph{evolutionary search}, which maintains a population of configurations and improves it through mutation and selection.

\subsubsection{Search range}

The search range constrains four discrete parameters: the human staffing level
$N \in \mathcal{N}$, the
release cadence $R \in \mathcal{R}$, the feature-freeze length $F_\mathrm{freeze} \in \mathcal{F}_\mathrm{freeze}$, and the bug-fixing share $f_\mathrm{fix} \in \mathcal{F}_\mathrm{fix}$.
The sets $\mathcal{N}$, $\mathcal{R}$, $\mathcal{F}_\mathrm{freeze}$, and $\mathcal{F}_\mathrm{fix}$ are
explicit finite sets, e.g., $\mathcal{N} = \{0.5, 1, 2\}$.

Once $N \in \mathcal{N}$ is fixed, the development composition $\mathbf{A}_\mathrm{dev}$ and the QA composition $\mathbf{A}_\mathrm{qa}$ must satisfy the structural constraints of Section~\ref{sec:model:team} and the human-capacity monitoring feasibility condition of Eq.~\eqref{eq:feas_monitoring}. However, there is a further constraint that the model implicitly imposes: the effort that every supervised role can contribute is restricted by the amount of work that reaches it, which is all derived from the feature arrival rate $r_\mathrm{arrival}$. While this is trivially the case for the implementation agent through Eqs.~\eqref{eq:impl_agents} and~\eqref{eq:impl}, the architecture work, bug fixing, test development, and documentation are also bounded by the inflow of work that reaches them (see Eqs.~\eqref{eq:arch_chg_done}, \eqref{eq:r_bugs_eff}, \eqref{eq:coverage}, and \eqref{eq:doc_cov}). Writing $\lambda_k$ for the work that reaches role $k$ in one development sprint, this gives us the following fixed-point upper bounds, expressed relative to the feature arrival rate $r_\mathrm{arrival}$:
\begin{align}
  \lambda_1 &= (1 + f_\mathrm{arch,chg})\,r_\mathrm{arrival}, &
  \lambda_2 &= f_\mathrm{arch,chg}\,r_\mathrm{arrival}, \nonumber\\
  \lambda_3 &= r_\mathrm{bugs}\,(2-\delta_\mathrm{min})\,r_\mathrm{arrival}, &
  \lambda_7 &= c_\mathrm{test,ceil}\,r_\mathrm{arrival}, \nonumber\\
  \lambda_8 &= r_\mathrm{arrival}.
  \label{eq:workload}
\end{align}
The workload is matched against the throughput $\eta_k$ an agent of role $k$ can achieve, accounting for the debt amplifier. We use its floor $\delta_\mathrm{min}$, i.e., the least favorable and therefore most permissive case:
\begin{equation}
  \begin{aligned}
    \eta_1 &= r_F\,\delta_\mathrm{min}, &
    \eta_2 &= r_\mathrm{arch,chg}, &
    \eta_3 &= r_\mathrm{fix}\,\delta_\mathrm{min}, \\
    \eta_7 &= r_\mathrm{test}, &
    \eta_8 &= r_\mathrm{doc}.
  \end{aligned}
  \label{eq:sat_rates}
\end{equation}

The two review services and the Refactoring Agent are not constrained by Eq.~\eqref{eq:sat_bound}, so instead of a search range we specify a sampling distribution for them below. The feature-freeze duration is additionally constrained by the release cadence: once $R$ is sampled, only values from $\mathcal{F}_\mathrm{freeze}$ with $0 \leq F_\mathrm{freeze} \leq R-1$ are admissible.

This gives us constraints
\begin{equation}
  a_{\mathrm{dev},k},\, a_{\mathrm{qa},k} \;\leq\; \bar{a}_k
    = \Bigl\lceil \sigma\,\frac{\lambda_k}{\eta_k} \Bigr\rceil,
  \qquad k \in \mathcal{K}_\mathrm{sup},
  \label{eq:sat_bound}
\end{equation}
where $\sigma \geq 1$ is a headroom factor that allows the search to explore configurations that exceed the steady-state workload, which means that the agent capacity can be overallocated to a certain degree to deal with load peaks that can occur due to the Poisson feature arrivals. We do not distinguish between the development and QA compositions here. Taken together with the feasible set of Section~\ref{sec:model:sosp}, this defines the region that the search explores:
\begin{equation}
  \mathcal{S} = \bigl\{\mathbf{c} \in \mathcal{X} \;:\;
    a_{\mathrm{dev},k} \leq \bar{a}_k \;\wedge\;
    a_{\mathrm{qa},k} \leq \bar{a}_k
    \;\;\forall k \in \mathcal{K}_\mathrm{sup} \bigr\} \subseteq \mathcal{X}.
  \label{eq:searched_region}
\end{equation}

\subsubsection{Random configuration sampling}
\label{sec:model:search:random}

The search needs a way to produce valid initial configurations. We sample these configurations as follows:
\begin{enumerate}
  \item Draw $N$, $R$, and $f_\mathrm{fix}$ uniformly from $\mathcal{N}$, $\mathcal{R}$, and $\mathcal{F}_\mathrm{fix}$, respectively.
  \item Draw $F_\mathrm{freeze}$ uniformly from the admissible values
        $\{f \in \mathcal{F}_\mathrm{freeze} : 0 \leq f \leq R-1\}$.
  \item Draw the supervised part of $\mathbf{A}_\mathrm{dev}$ role by role, independently and
        uniformly within the throughput saturation bound of Eq.~\eqref{eq:sat_bound}, i.e.,
        $a_{\mathrm{dev},1}$ from $\{1, \ldots, \bar{a}_1\}$ so that
        $a_{\mathrm{dev},1} \geq 1$, and $a_{\mathrm{dev},k}$ from
        $\{0, \ldots, \bar{a}_k\}$ for the remaining $k \in \mathcal{K}_\mathrm{sup}$.
  \item Draw the supervised part of $\mathbf{A}_\mathrm{qa}$ in the same way, i.e.,
        $a_{\mathrm{qa},k}$ uniformly from $\{0, \ldots, \bar{a}_k\}$ for
        $k \in \{3,7,8\}$, and $a_{\mathrm{qa},1} = a_{\mathrm{qa},2} = 0$. If
        $F_\mathrm{freeze} = 0$ was drawn in step~2, then $\mathbf{A}_\mathrm{qa} = \mathbf{0}$
        by Section~\ref{sec:model:team} and this step is skipped, as is the QA half of
        step~5.
  \item Draw the review service counts $a_{\mathrm{dev},5}$, $a_{\mathrm{dev},6}$ and,
        when $F_\mathrm{freeze} \geq 1$, $a_{\mathrm{qa},5}$, $a_{\mathrm{qa},6}$
        independently and uniformly from
        $\{0, 1, \ldots, \lfloor r_\mathrm{arrival} \rfloor\}$, i.e., we allow up to one instance
        of each service per feature that arrives in a sprint.
  \item Draw the Refactoring Agent counts $a_{\mathrm{dev},4}$ and, when $F_\mathrm{freeze} \geq 1$,
        $a_{\mathrm{qa},4}$ independently and uniformly from 
        $\{0, 1, \ldots, \bar{a}_4\}$ with $\bar{a}_4 = \lfloor r_\mathrm{arrival}/2 \rfloor$,
        i.e., we allow up to one Refactoring Agent per two features that arrive in a sprint.\footnote{While this is only based on features, the number of bugs is limited by the number of features, which also limits the refactorings.}
  \item Check the candidate against the human-capacity feasibility condition of
        Eq.~\eqref{eq:feas_monitoring}. A candidate that violates it is recorded as
        infeasible and not simulated.
\end{enumerate}

Note that the sampling does not require that the throughput constraint for the value achieved is fulfilled. 

\subsubsection{Constraint handling}

The advantage of not discarding solutions that violate the throughput constraint is that we can approach the admissible region from the outside, possibly finding better configurations. Still, this means we must be able to compare two configurations of which one, or both, violate Eq.~\eqref{eq:throughput_constraint}. We adopt a feasibility rule from \cite{deb2000efficient} to resolve this:
\begin{itemize}
  \item If both configurations violate the throughput constraint, the one with the smaller constraint violation, i.e., the higher value, is preferred.
  \item A configuration that satisfies the throughput constraint is always preferred to one that does not.
  \item Two configurations that both satisfy the throughput constraint are compared by their cost, with the lower cost preferred.
\end{itemize}
Thus, the search first looks for admissible configurations, and then among those for the cheapest one. We use the notation $\mathbf{c} \prec \mathbf{c}'$ to denote that configuration $\mathbf{c}$ is preferred to configuration $\mathbf{c}'$ according to the above rules.

\subsubsection{Evolutionary search}
\label{sec:model:search:evo}

Evolutionary search is a metaheuristic that iteratively generates and evaluates candidate solutions. It maintains a population of candidates and favors better candidates when generating new ones. In our case, we use a simple mutation-based algorithm that creates new solutions by modifying existing ones.

\paragraph{Initialization.}
The evolutionary search starts with an initial pool $\mathcal{P}_0$ of $n_\mathrm{init}$ configurations, drawn with the random sampling procedure of Section~\ref{sec:model:search:random}, so that the initial pool is diverse and covers a broad area of the search space. Each configuration in the initial pool is evaluated with $K'$ Monte Carlo replications.

\paragraph{Mutation operator.}
The mutation operator receives a parent configuration
$\mathbf{c} = (N, \mathbf{A}_\mathrm{dev}, \mathbf{A}_\mathrm{qa}, R, F_\mathrm{freeze}, f_\mathrm{fix})$
and produces a child by perturbing exactly one randomly chosen parameter.
$R$ is never mutated, i.e., the release cadence is only sampled when generating the initial population.
The parameter to perturb is selected uniformly at random from five choices, or from the
first four when $F_\mathrm{freeze} = 0$:
\begin{itemize}
  \item \emph{Mutate $N$:} replace the current value with an adjacent value in the
        discrete set $\mathcal{N}$ (the next smaller or larger element).
  \item \emph{Mutate $F_\mathrm{freeze}$:} replace the current value with an adjacent value in
        the admissible set $\{f \in \mathcal{F}_\mathrm{freeze} : 0 \leq f \leq R-1\}$.
  \item \emph{Mutate $f_\mathrm{fix}$:} replace the current value with an adjacent value in
        $\mathcal{F}_\mathrm{fix}$.
  \item \emph{Mutate $\mathbf{A}_\mathrm{dev}$:} apply one elementary operation drawn uniformly from:
        \begin{itemize}
          \item \emph{Add}: increment one randomly chosen supervised role by~1, provided it
                stays at or below its saturation bound $\bar{a}_k$.
          \item \emph{Remove}: decrement one randomly chosen non-empty supervised role by~1.
          \item \emph{Move}: select a random non-empty supervised donor role and transfer one agent to a randomly chosen
                supervised recipient role that is still below its saturation bound $\bar{a}_k$.
          \item \emph{Adjust service}: change one of the three autonomous roles
                $\mathcal{K}_\mathrm{aut} = \{4,5,6\}$ by $\pm m$, where the sign is drawn uniformly
                and the step size is $m = 2^j$ with $j$ drawn uniformly from
                $\{0, 1, \ldots, \lfloor \log_2 s_k \rfloor\}$ and $s_k$ the sampling range of the
                role in steps~5 and~6 of Section~\ref{sec:model:search:random}. The result is floored at zero and, for $k=4$, capped at $\bar{a}_4$. 
        \end{itemize}
  \item \emph{Mutate $\mathbf{A}_\mathrm{qa}$:} works the same as for $\mathbf{A}_\mathrm{dev}$, except that the implementation and architecture roles ($k = 1, 2$) remain fixed at zero. This choice is only available when $F_\mathrm{freeze} \geq 1$, i.e., the QA composition can only be modified if it is active.
\end{itemize}
The adjustment of the three autonomous roles uses the exponential set of step size options to be able to navigate the large range of possible counts efficiently. 

If a mutation attempt violates the structural constraints of Section~\ref{sec:model:team} or
leaves the parent unchanged (e.g., because the chosen role is already at its
saturation bound, or because a randomly chosen operation
is a structural no-op), the attempt is discarded and a new mutation type is drawn.
Up to ten attempts are made. If none of them yields a distinct valid child, the parent is
returned unchanged.

\paragraph{Generation loop.} The generation loop iteratively produces new configurations by mutating selected parents from the current pool of configurations. The loop is repeated for $g=1,\ldots,G$ generations. In each generation, the following steps are performed:
\begin{enumerate}
  \item Rank the previously generated population $\mathcal{P}_{g-1}$ by the preference relation $\prec$ and let the elite $\mathcal{E}_{g-1}$ be its $n_\mathrm{elite}$ best configurations.
  \item Sample $n_\mathrm{off}$ parent configurations. 80\% of the parents are drawn uniformly from $\mathcal{E}_{g-1}$ to search for improvements near the best configurations found so far. 20\% of the parents are drawn uniformly from $\mathcal{P}_{g-1} \setminus \mathcal{E}_{g-1}$ to enable exploration of the search space in other regions.
  \item Apply the mutation operator to each parent, creating the child population $\mathcal{P}_{g,\mathrm{child}}$.
  \item Create the new population $\mathcal{P}_g$ as the union of the child population and the current population: $\mathcal{P}_g = \mathcal{P}_{g-1} \cup \mathcal{P}_{g,\mathrm{child}}$.
\end{enumerate}

The mixture of drawing most mutations from the elite and some from the rest of the population is designed to balance exploitation and exploration: the former focuses the search on improving the best-known configurations, while the latter allows for discovering new regions of the search space that may contain better configurations.

\paragraph{Final selection.}
At the conclusion of the generation loop, $\mathbf{c}^\star$ and the indistinguishable set $\mathcal{I}^\star$ are determined as described in Section~\ref{sec:model:selection} over all sampled configurations $\mathcal{P}_G$, including the re-simulation of the preliminary $\mathcal{I}^\star$ with $K$ replications.

\subsection{Limitations of the simulation's scope}
\label{sec:model:limitations}

The simulation model is built with the intent to be able to analyze cost trade-offs and understand team compositions. The aspects that we simulate were added because we believe they are key factors to understand these trade-offs. However, our model is only an abstraction and makes simplifying assumptions about software development:
\begin{itemize}
  \item We use fixed efforts and costs for all features and bugs. An alternative would be to sample required efforts and costs, similar to how we sample feature arrival rates. However, the expected rates should converge to the average that we use, if simulations run sufficiently long. Thus, this mainly matters for estimating short-term uncertainty and not long-term effects, which is why we opted against this additional complexity and only randomize the number of features that are arriving and the bug introduction rates. 
  \item Related to the above, we also do not increase development costs for larger systems, e.g., as a factor of features already implemented. This simplification makes the estimation of cost parameters easier, but means that we have to assume an average cost over time, which might underestimate costs for larger systems over time. However, assuming that this affects human costs and agent costs similarly, this should only have a small effect on our results. 
  \item Our simplification that human effort is debt-independent simplifies the model greatly, because otherwise we would also have to model test generation and technical debt reduction for human developers, including a possibly complex model of how human developers allocate their time between these activities and the development activities feature implementation and bug fixes. We believe that this simplification is reasonable, because the human effort estimates that we use already include the work of coping with the complexity of an existing code base. However, this does not paint a fully realistic picture of how technical debt also slows down human developers over time. 
  \item We implicitly assume that the re-inspection of the code during a feature freeze is independent of the inspections conducted when the features were implemented, i.e., bugs missed by the code review or security audit before have the same chance of being detected during the freeze as newly injected bugs. This is an optimistic assumption, since a reviewer that missed a bug once is likely to miss it again. However, given the stochastic nature of LLMs and the possibility to use different LLMs for different inspection phases, this assumption is not completely unreasonable. Furthermore, the alternative would be to estimate how much the detection probability is reduced with repeated reviews, for which, to the best of our knowledge, no data is available.
\end{itemize}

% ============================================================
\section{Experimental setup}
\label{sec:setup}
% ============================================================

% ---- 4.1 ----
\subsection{Human-only baseline}
\label{sec:setup:baseline}

An important baseline to understand the impact of agents on software development is to compare mixed teams against a baseline with only human developers. For this, we set all agent counts to zero, i.e., $\mathbf{A}_\mathrm{dev} = \mathbf{A}_\mathrm{qa} = \mathbf{0}$, and let the human developers implement all features and fix all bugs themselves. The other model parameters are still optimized, i.e., the number of FTE developers $N$, release cadence $R$, feature-freeze length $F_\mathrm{freeze}$, and bug-fixing share $f_\mathrm{fix}$ are still decision variables. The costs remain the same, except that the agent costs are always zero. 

However, we need to slightly modify the simulation model to support this, since we modeled the debt removal and bug detection as purely agentic tasks. We use a simplified model without debt dynamics for the human-only case, built on the assumption that some debt-removal and architecture updates are performed as part of the normal development process by the human developers. For bug detection, we reserve a share $f_\mathrm{qa} \in [0,1)$ of the developers' net capacity for quality assurance, which the baseline optimizes as an additional decision variable:
\begin{equation}
  H_\mathrm{after\_qa}(t) = (1 - f_\mathrm{qa})\,H_\mathrm{net}.
  \label{eq:hafter_qa}
\end{equation}
The time $H_\mathrm{after\_qa}(t)$ is then used in place of $H_\mathrm{after\_esc}(t)$ in the implementation and bug-fixing equations. From a modeling perspective, this means that the developers spend this time on quality assurance instead of supervising agents.

The detection itself is implemented as a variant of the agent-driven detection which replaces the bug detection in Step~5. We match the reserved capacity $f_\mathrm{qa}\,N\,H_\mathrm{net}$ against the review effort of the sprint, which is $h_\mathrm{qa}$ hours for each of the $v(t)$ features under inspection (Eq.~\eqref{eq:inspected}). These are the features implemented in a development sprint and all features of the current release cycle in a feature-freeze sprint. This gives us the review coverage
\begin{equation}
  \mathrm{cov\_qa}(t) = \min\!\Bigl(1,\;
    \frac{f_\mathrm{qa}\,N\,H_\mathrm{net}}
         {h_\mathrm{qa}\,\max(1,\,v(t))}\Bigr),
  \label{eq:cov_qa}
\end{equation}
i.e., the fraction of the inspected features that can be reviewed in full, which is similar to the volume factor $\Phi$ we use for agentic review in Eq.~\eqref{eq:volume_factor}. With a human bug detection capability factor $d_\mathrm{qa}$, the rate of bug detection is then
\begin{equation}
  \mathrm{QA}_x(t) \sim \mathrm{Binomial}\!\bigl(P_x(t),\; \mathrm{cov\_qa}(t)\,d_\mathrm{qa}\bigr),
    \qquad x \in \{\mathrm{min,maj,sec}\}.
  \label{eq:qa_human}
\end{equation}
The detected bugs are then removed from the hidden queues and moved to the known queues, analogous to the agentic detection. 

% ---- 4.2 ----
\subsection{Scenarios}
\label{sec:setup:scenarios}

Five orthogonal aspects influence our simulation, each governed by a distinct group of scenario parameters that we define in the following subsections: human developers, project dynamics, debt dynamics, agent costs and throughput, and bug impact. Each group contains multiple parameters that govern different facets of that aspect. For each of these aspects, we define three levels of parameters, corresponding to a low, medium, and high setting, and additionally two levels that stress the simulation beyond the observed data: a very high level for the project dynamics and a pessimistic level for the agent costs and throughput. For example, agent capabilities can be lower or higher, varying the correctness and costs of agentic actions. Orthogonally, developers may be junior, mid-level, or senior, resulting in different capabilities and salaries. 
Since the groups are orthogonal, we can freely combine all five groups into a diverse set of scenarios. 

\subsubsection{Human developer parameters}

The scenario parameters for human developers are summarized in Table~\ref{tab:scenario:human}.
We assume that a full-time developer works a 40-hour week, i.e., $H_\mathrm{capacity}$ is stated per FTE and a staffing level of $N=0.25$ corresponds to a ten-hour week. We assume that the communication overhead is one hour per team member. The salaries are estimated based on average salaries reported by Indeed~\citep{indeed2026salaries}. We round low, average and high salaries to the next \$50k, to account for the employer overhead costs.

While there is no data yet on how much time developers require to monitor agents, early results indicate that there is a fairly low interruption frequency~\citep{baumann2026swechat}, so we set the monitoring overhead at a modest 2.5 hours per week, i.e., 30 minutes per day. The work capacity $H_\mathrm{capacity}$, the communication overhead $c_\mathrm{comm}$, and the monitoring overhead $H_\mathrm{mon,overhead}$ are independent of developer skill and are therefore the same across all three tiers. There are also no estimates yet of how much time developers require to resolve an agent escalation. It is reasonable to assume that these are non-trivial problems that require developers to get involved with the code base, so we assume that the time required is similar to that required to implement a feature (estimate see below).

We use the review of architectural decision records (ADRs) to ground the architecture-review time within our model. Amazon Web Services estimates 10--15 minutes to read the document~\citep{amazonArchitecturalDecision}. The process is assumed to be iterative and done by multiple developers. Assuming that updating the ADR based on comments requires 45 minutes, one iteration would require 1 hour. With 4 iterations, this would require a total of 4 hours, which we use for the medium tier. We assume that less capable developers require twice as long and highly capable developers are twice as fast, resulting in 8 hours for the low tier, and 2 hours for the high tier.

To have a comparable baseline for human effort, we assume the time required to solve SWE-bench Verified coding tasks, which are designed to be representative of real-world coding tasks by basing them on issues reported in open-source repositories~\citep{openai2026swebenchverified,jimenez2024swe}. We use METR's estimate for the human baseline of the time required to solve substantial issues within the 1--4 hour time bucket of the benchmark, which estimates the time at 2 hours~\citep{kwa2026measuringaiabilitycomplete}. Since this seems reasonable, we use this for the medium tier. We use the pattern that less capable developers require twice as long as the medium tier and highly capable developers are twice as fast, resulting in 4 hours for the low tier, and 1 hour for the high tier. Since METR found that the $<15$ minute time bucket underestimates the time required to solve tasks, we instead use the 15 to 60 minute time bucket to estimate the time required to fix bugs in a similar manner, i.e., 30 minutes for the medium tier, 60 minutes for the low tier, and 15 minutes for the high tier.

To have comparable bug introduction rates between humans and the agents we later define, we ground the bug introduction rate of humans in research by \cite{coderabbit2025aicode}, who report that AI introduces issues at 1.7 times the rate of humans. Since this number is one year old at the time of writing, we assume that this reflects the capability of the low agentic tier, which translates into a bug introduction rate of 0.12 for humans. Since junior developers are often compared to AI, we use the same bug introduction rate as the AI, i.e., 0.20 for the low tier. For the more capable high tier developers, we use a rate of 0.08, using the same ratio of performance differences as before. 

Finally, we need to estimate the human quality assurance capability for the human-only baseline. We use a range of values based on a large study on code-review effectiveness by \cite{cohen2006best}: they found that the review of 200--400 lines requires 60 to 90 minutes and leads to a discovery of 70--90\% of defects. Since we assume that a feature has 107 lines (Section~\ref{sec:scenario:agentcosts}), the review time for a feature is between 32 and 48 minutes. For the time per review $h_\mathrm{qa}$, we use the fastest time of 32 minutes to ground the values for the high developer tier, the slowest time of 48 minutes for the low tier and 37 minutes for the medium tier. For the detection rates $d_\mathrm{qa}$, we use the midpoint from \cite{cohen2006best} for all tiers, i.e., 0.8, since we already encode the difference between tiers in the time required.

\begin{table}[t]
\caption{Human developer parameters. Low represents a junior developer, medium a mid-level developer, and high a senior developer.}
\label{tab:scenario:human}
\footnotesize
\begin{tabularx}{\linewidth}{p{1.8cm}rrrX}
\toprule
Symbol & Low & Medium & High & Source \\
\midrule
$H_\mathrm{capacity}$         & 40.0 & 40.0 & 40.0 & One 40h work-week per FTE. \\[2pt]
$c_\mathrm{comm}$             & 1.0  & 1.0  & 1.0  & Assumed. \\[2pt]
$H_\mathrm{mon,overhead}$     & 2.5  & 2.5  & 2.5  & Assumed based on \cite{baumann2026swechat}. \\[2pt]
$H_\mathrm{esc\_time}$        & 4.0  & 2.0  & 1.0  & Assumed to be similar to feature implementation time. \\[2pt]
$H_\mathrm{review\_time}$     & 8.0  & 4.0  & 2.0  & Extrapolated from \cite{amazonArchitecturalDecision}. \\[2pt]
$h_\mathrm{feature}$          & 4.0  & 2.0  & 1.0  & Based on SWE-bench Verified, 1--4h bucket \citep{openai2026swebenchverified,kwa2026measuringaiabilitycomplete}. \\[2pt]
$h_\mathrm{fix}$              & 1.0 & 0.5 & 0.25 & Based on SWE-bench Verified, 0.25--1h bucket \citep{openai2026swebenchverified,kwa2026measuringaiabilitycomplete}. \\[2pt]
$r_\mathrm{bugs,human}$   & 0.20 & 0.12 & 0.08 & \cite{coderabbit2025aicode}. \\
$h_\mathrm{qa}$ & 0.80 & 0.62 & 0.53 & Based on \cite{cohen2006best}. \\
$d_\mathrm{qa}$               & 0.8 & 0.8 & 0.8 & \cite{cohen2006best}. \\
$c_\mathrm{salary}$ & 100{,}000 & 150{,}000 & 250{,}000 & \cite{indeed2026salaries}. \\
\bottomrule
\end{tabularx}
\end{table}

\subsubsection{Project dynamics parameters}

To understand the project dynamics, we first collected data from all projects with at least 100 issues and at least 10 stars using SEART's GitHub Search tool~\citep{dabic2021data}. We then filtered these projects such that they have an average activity of at least 10 commits per week, i.e., they were consistently active during their lifetime. From the resulting projects, we considered the 1{,}000 projects with the most stars, i.e., that are popular and actively engaged by the community. We then compute the number of issues per week and use the 5th percentile for the low tier, the 50th percentile for the medium tier, and the 95th percentile for the high tier for the number of issues per week. We note that these estimates are only valid for open-source projects and may not be representative of closed-source projects. Moreover, since a quarter FTE supported by agents can already deliver the throughput of the high tier, this tier does not stress the simulation to assess extreme-velocity scenarios. Therefore, we add a very high tier as well, which has 1{,}000 issues per week. Although unrealistically high, this tier shows if and how such extreme scales could be managed by a single team. We note that this tier also stresses the architecture pipeline beyond what human review can absorb: at $f_\mathrm{arch,chg} = 0.13$ it generates around 130 change requests per sprint, which only the fastest reviewers of the high human tier can keep up with even at the largest staffing level we consider. The resulting parameters are summarized in Table~\ref{tab:scenario:project}.

To understand how often issues affect architecture, we use data from \cite{shahbazian2018toward}, who used automated methods to determine whether the change triggered by an issue caused architectural changes. Depending on the method used, the proportion ranged from 13\% of issues to 26\% of issues. Since the automated approach possibly overestimates the proportion by counting purely structural changes, we use the lower bound of 13\%. We derive the data for the distribution of the severity of bugs from data collected by \cite{ohira2015highimpact}. They distinguish between critical, blocker, and major issues (which we map to major bugs), and minor and trivial issues (which we map to minor bugs). Orthogonally, they determine which of these bugs are security related, which we use to infer the ratio of security bugs. We exclude one of the collected projects (Ambari) from the estimation of the ratio of major bugs, since it did not mark issues as minor.
These numbers are independent of the number of issues per week and therefore the same across all four tiers.

\begin{table}[t]
\caption{Project dynamics parameters guiding the feature arrival and bug severity distributions.}
\label{tab:scenario:project}
\footnotesize
\begin{tabularx}{\linewidth}{p{1.8cm}rrrrX}
\toprule
Symbol & Low & Medium & High & Very High & Source \\
\midrule
$r_\mathrm{arrival}$  & 3  & 12  & 75  & 1{,}000 & Collected data. \\[2pt]
$f_\mathrm{arch,chg}$  & 0.13  & 0.13  & 0.13  & 0.13 & \cite{shahbazian2018toward}. \\[2pt]
$\theta_\mathrm{maj}$ & 0.44 & 0.44 & 0.44 & 0.44 & \cite{ohira2015highimpact}. \\[2pt]
$\theta_\mathrm{sec}$ & 0.07 & 0.07 & 0.07 & 0.07 & \cite{ohira2015highimpact}. \\
\bottomrule
\end{tabularx}
\end{table}

\subsubsection{Debt dynamics parameters}

The debt dynamics parameters are summarized in Table~\ref{tab:scenario:debt}. There is a limited amount of literature on the impact of documentation and the studies are typically conducted with small sets of developers, limiting their ability to find significant effects. \cite{dzidek2008uml} found an increase in maintenance time of 40\% when no UML documentation was made available based on a group of 20 developers. In a recent eye tracking experiment using 12 (mostly junior) developers, \cite{abdelsalam2026comments} found an increase in program comprehension time of up to 34\% based on code comments, but this effect depended on the complexity of the code and comprehension time was even reduced for simpler code snippets. Based on these studies, we set the low tier to 40\% (the effect found by \cite{dzidek2008uml}), the medium tier to 34\% (the maximum effect found by \cite{abdelsalam2026comments}), and the high tier to 0\% (no effect, due to the lack of significance of the results).

We estimate the addition of technical debt based on data from \cite{he2026cursordebt}, who found that AI-assisted coding increases code complexity and static analysis warnings. We note that this is particular to AI-assisted development as the technical debt otherwise seems to be decreasing over time~\citep{trautsch2020longitudinal}. Since the average treatment effect found that AI-assisted development leads to 28\% more lines, but 30\% more static analysis warnings, we assume that the debt increases by at least 2\% for every feature added. Since \cite{he2026cursordebt} also found that the complexity increases by 41\% due to the use of AI, we assume that the upper bound for the debt increase is 13\% per feature added. We therefore set the low tier to 2\%, the medium tier to 6\% and the high tier to 13\% per feature added. Since \cite{trautsch2023automated} found that there is no detectable difference between the static analysis warnings added by bug fixing commits and other commits, we use the same values for the debt increase due to bug fixing. 

We set the values for the debt amplifier floor $\delta_\mathrm{min}$, i.e., the fraction of the nominal throughput that is retained at maximal debt, based on a developer survey from \cite{besker2018technical}. They found that developers spend on average 23\% of their time dealing with technical debt with a standard deviation of 17\%. We interpret this share $s$ of lost time as the slow-down at maximal debt and hence set $\delta_\mathrm{min} = 1 - s$. We therefore set the low value at $0.94$ (i.e., a slow-down of 6\%, one standard deviation below the average), the medium value at $0.77$ (a slow-down of 23\%, the average), and the high value at $0.60$ (i.e., a slow-down of 40\%, one standard deviation above the average). \cite{martini2017interest} found that technical debt can be contagious, which supports a non-linear effect of the technical debt. Since there is no data on how strong this non-linearity is, we assume a slightly super-linear effect with $\delta_\mathrm{exp} = 1.5$ across all scenarios.

\begin{table}[t]
\caption{Debt dynamics parameters across scenarios. Note that $\delta_\mathrm{min}$ is the throughput fraction \emph{retained} at maximal debt, i.e., lower values indicate a stronger debt penalty.}
\label{tab:scenario:debt}
\footnotesize
\begin{tabularx}{\linewidth}{p{1.8cm}rrrX}
\toprule
Symbol & Low & Medium & High & Source \\
\midrule
$\delta_\mathrm{impl}$ & 0.02 & 0.06 & 0.13 & \cite{he2026cursordebt}. \\[2pt]
$\delta_\mathrm{fix}$  & 0.02 & 0.06 & 0.13 & \cite{he2026cursordebt}. \\[2pt]
$\delta_\mathrm{min}$  & 0.94  & 0.77  & 0.60  & \cite{besker2018technical}. \\[2pt]
$\delta_\mathrm{exp}$  & 1.5   & 1.5   & 1.5   & Assumed based on contagiousness of debt~\citep{martini2017interest}. \\[2pt]
$\Gamma_\mathrm{doc}$  & 0.40  & 0.34  & 0.0  & \cite{dzidek2008uml, abdelsalam2026comments}. \\
\bottomrule
\end{tabularx}
\end{table}

\subsubsection{Agent costs and throughput parameters}
\label{sec:scenario:agentcosts}

The agent cost parameters are summarized in Table~\ref{tab:scenario:agents}. While there is no direct scientific evidence for the exact costs per feature implementation, we assume that implementing a feature requires 100 tool calls, 2M cumulative input tokens, and 40K output tokens~\citep{morph2026aicodingcosts}. For bug fixes, we assume 25 tool calls, 400K cumulative input tokens, and 10K output tokens~\citep{morph2026aicodingcosts}. The costs for the other activities are grounded on these two costs:
\begin{itemize}
  \item Code review is assumed to be comparable to bug fixing, since similar analysis needs to be performed to understand the code and its implications. Security audits are more thorough and assumed to be ten times more expensive than normal code review.
  \item Architecture and design changes possibly require an analysis of the whole code base to understand the implications. Therefore, we assume that these carry high costs, comparable to the cost of implementing five features.
  \item Refactoring is assumed to be comparable to implementing a feature, since it requires a similar amount of analysis and code changes. Since we assume that refactorings are required to account for the debt introduced by feature additions and bug fixes, we scale the parameters with respect to this and assume that each Refactoring Agent can reduce the debt introduced for a single feature or bug fix. Consequently, we set the rates based on $\delta_\mathrm{impl}$ and $\delta_\mathrm{fix}$. We assume that the agent tiers' capabilities are comparable to the debt dynamics, giving us a debt reduction of 0.02 for the low tier, 0.06 for the medium tier, and 0.13 for the high tier. We then set the cost of refactoring to be the same as the cost of implementing a feature, since it requires code analysis and code changes that might affect multiple files. 
  \item Writing tests and documentation are assumed to be comparable to bug fixing, since they require a similar analysis of a local context.
\end{itemize}
We further ground the pricing based on the API pricing of Claude Sonnet~5, Opus~5, and Fable~5 for low, medium, and high, respectively.

We model the agent throughput on the latency of the models, i.e., the time to solve a task. Since a concrete latency is hard to define as this depends not only on model size, but also on the current load of cloud providers, we do not base this estimate on single reported latency values. As for the other tasks, we ground this in SWE-bench, where the reference solutions require 107 lines of code on average~\citep{scaleSWEBenchPublic}. Based on data from \cite{baumann2026swechat}, autonomous coding required 12.6 minutes on average per 100 commit lines. Assuming that agents only run during work hours, this would result in 178 features per sprint. Since this measurement is not specific to a model, we use a single throughput value across all tiers and derive the rates of the remaining roles from it with the same ratios in every tier, as reported in Table~\ref{tab:scenario:agents}.

We estimate the agentic bug introduction rate based on the SWE-bench Verified benchmark score, assuming that each failure in the benchmark results in a bug in the code base. We collect the data for Claude Sonnet~5, Opus~5, and Fable~5 analogously to the cost estimations. The values are taken from the vals.ai leaderboard~\citep{valsai2026swebench}, i.e., we do not use Anthropic's estimates but those of independent benchmark executions.\footnote{On the leaderboard, Opus~5 has a marginally higher score than Fable~5. We attribute this to a measurement artifact and use the best-performing value for the high tier and the second-best for the medium tier for internal consistency.} While we are aware that these results likely overestimate the actual benchmark performance~\citep{liang2026swe}, we nevertheless use this as an (optimistic) estimate for the capabilities of the agents.

We use CyberBench~\citep{liu2024cyberbench} to estimate the security bug detection rate. While the benchmark is not a perfect fit, since it also includes the correction of the bug, it still gives us a good baseline for the detection rate. For other benchmarks, like SEC-Bench Pro, we could not find data for recent models that are not self-reported by the model providers. Due to the model guardrails, we also cannot use the performance of the Claude~5 models here and instead use the GPT-5.6 models (Luna, Sol) and the scores reported by vals.ai~\citep{valsai2026swebench} for the medium and high tiers. Since no data is available for GPT-5.6 Terra either, we assume that the difference between Terra and Luna is the same as between Luna and Sol. We estimate the capability to generate tests based on data from \cite{huang2025benchmarking}, who developed a benchmark with unit tests for functions that were not part of the training. They found that test generation capability strongly correlates with coding ability, but models scored lower on the test generation benchmark, with differences of at least 0.2. Since no similar benchmark exists for the latest models, we assume that the trend holds and use the same difference here, i.e., the SWE-bench scores reduced by 0.2.

We use SWE-Review~\citep{wang2026swe} to estimate how good language models are at code review. Within this benchmark, Claude Opus 4.6 was able to correctly judge pull requests for between 75\% and 89\% of the cases. We use these as the low and high tier, respectively, and use the midpoint 0.82 for the medium tier. To estimate the ceiling of what testing can achieve, we use the results of \cite{SCHWARTZ2018106}, who found that even for highly covered projects, 46\% of mutants could not be detected by the test suite, giving a ceiling of 0.54, which we use as the medium estimate. We assume that less capable agents are worse at testing, while more capable agents are better, resulting in a low tier of 48\% and a high tier of 60\%.

Finally, we estimate the escalation rate based on telemetry data from \cite{baumann2026swechat}: in their study, the agents escalated between 1\% and 3\% of the time, which we use as the high and low tier, respectively, since less capable agents are expected to escalate more often. We use the midpoint 0.02 for the medium tier.

All of the estimates above are optimistic, both regarding the capabilities and the costs of agents. On the capability side, the benchmark scores likely overstate what agents achieve on real code bases, as we already noted for the bug introduction rate. On the cost side, the prices are those of the current API tariffs applied to a fixed assumption about how many tokens a work item consumes, so they account neither for price increases nor for agents that need several attempts to solve a task. We therefore add a fourth, pessimistic tier for this group. Like the very high project tier, this tier is not based on observed data. It tests whether our conclusions about agent-heavy teams still hold if our assumptions are too optimistic. The pessimistic tier combines the capability of the low tier with ten times the cost of the high tier. Hence, all parameters of the pessimistic tier except $c_\mathrm{token}$ are identical to those of the low tier.

\begin{table}[t]
\caption{Agent cost, capability, and effort parameters across scenarios, grouped by role. $c_\mathrm{token}$ is USD per unit of the role's throughput rate $r_k$.}
\label{tab:scenario:agents}
\footnotesize
\setlength{\tabcolsep}{3pt}
\begin{tabularx}{\linewidth}{p{2.2cm}p{1.5cm}rrrrX}
\toprule
Role & Parameter & Low & Medium & High & Pess. & Source \\
\midrule
\multirow{3}{2.2cm}{1 -- Feature Impl.}
  & $c_\mathrm{token}$     & 2.28 & 3.80 & 7.60 & 76.00 & \cite{morph2026aicodingcosts} \\
  & $r_F$                  & 178  & 178  & 178  & 178   & Based on 12.6\,min per 100 lines of code~\citep{baumann2026swechat,scaleSWEBenchPublic} \\
  & $r_\mathrm{bugs}$      & 0.20 & 0.05 & 0.03 & 0.20  & \cite{valsai2026swebench} \\
\midrule
\multirow{2}{2.2cm}{2 -- Architecture \& Design}
  & $c_\mathrm{token}$     & 11.40 & 19.00 & 38.00 & 380.00 & 5 times the cost of feature implementation \\
  & $r_\mathrm{arch,chg}$  & 35    & 35    & 35    & 35     & Five times slower than feature implementation \\
\midrule
\multirow{2}{2.2cm}{3 -- Bug Fixing}
  & $c_\mathrm{token}$     & 0.54 & 0.90 & 1.80 & 18.00 & \cite{morph2026aicodingcosts} \\
  & $r_\mathrm{fix}$       & 747  & 747  & 747  & 747   & 4.2 times faster than feature implementation, based on cost estimate \\
\midrule
\multirow{2}{2.2cm}{4 -- Refactoring}
  & $c_\mathrm{token}$     & 2.28 & 3.80 & 7.60 & 76.00 & One feature implementation per sprint \\
  & $r_\mathrm{refactor}$  & 0.02 & 0.06 & 0.13 & 0.02  & Can remove debt of one feature/bug fix \\
\midrule
\multirow{3}{2.2cm}{5 -- Code Review}
  & $c_\mathrm{token}$     & 0.54 & 0.90 & 1.80 & 18.00 & Same as bug fixing \\
  & $d_\mathrm{CR}$        & 0.75 & 0.82 & 0.89 & 0.75  & \cite{wang2026swe} \\
  & $\alpha_\mathrm{size}$ & 0.10 & 0.10 & 0.10 & 0.10  & Assumed, with one feature per detection agent as reference volume \\
\midrule
\multirow{2}{2.2cm}{6 -- Security Audit}
  & $c_\mathrm{token}$     & 5.40 & 9.00 & 18.00 & 180.00 & 10 times the cost of code review \\
  & $d_\mathrm{SA}$        & 0.78 & 0.83 & 0.88  & 0.78   & \cite{liu2024cyberbench}, scores from \cite{valsai2026swebench} \\
\midrule
\multirow{4}{2.2cm}{7 -- Test Development}
  & $c_\mathrm{token}$     & 0.54 & 0.90 & 1.80 & 18.00 & Same as bug fixing \\
  & $r_\mathrm{test}$      & 747  & 747  & 747  & 747   & Same as bug fixing \\
  & $d_\mathrm{test}$      & 0.60 & 0.75 & 0.77 & 0.60  & Based on \cite{huang2025benchmarking} and \cite{valsai2026swebench} \\
  & $c_\mathrm{test,ceil}$ & 0.48 & 0.54 & 0.60 & 0.48  & \cite{SCHWARTZ2018106} \\
\midrule
\multirow{2}{2.2cm}{8 -- Dev.\ Documentation}
  & $c_\mathrm{token}$     & 0.54 & 0.90 & 1.80 & 18.00 & Same as bug fixing \\
  & $r_\mathrm{doc}$       & 747  & 747  & 747  & 747   & Same as bug fixing \\
\midrule
All roles & $\rho_\mathrm{esc}$ & 0.03 & 0.02 & 0.01 & 0.03 & \cite{baumann2026swechat} \\
\bottomrule
\end{tabularx}
\end{table}

\subsubsection{Bug impact parameters}
\label{sec:scenario:bugimpact}

The weights $\omega_\mathrm{min}$, $\omega_\mathrm{maj}$, and $\omega_\mathrm{sec}$ are the expected cost in dollars that a minor, major, or security bug causes once it escapes to production and affects real users. We ground them in real-world estimates of such costs, distinguishing three system contexts that match our low, medium, and high tiers: a simple web application (e.g., a small SaaS product), an enterprise system with a large user base and a formal incident-response process, and a safety-critical system such as a medical device. 

We estimate the cost of a minor bug as the direct engineering cost of repairing it after release, which we assume to be comparable to implementing a feature, i.e., two hours of developer time. We compute this using a fully-loaded hourly rate typical for the respective system context (assuming a 2{,}080-hour work year), giving \$100, \$150, and \$250 for the low, medium, and high tier, respectively. We therefore assume that such bugs have no external incident costs, because they are only a nuisance and do not affect functionality.

For major bugs, i.e., bugs that break or degrade a feature in production for a non-trivial period, we ground our estimate in reported downtime costs. For the low tier, ITIC's 2024 downtime survey reports that small and mid-sized businesses commonly estimate the cost of an hour of downtime at around \$25{,}000~\citep{itic2026downtime}. For the medium tier, industry estimates put the cost of an hour of downtime for a critical enterprise application at over \$300{,}000~\citep{cloudqa2026bugcost}, which is consistent with ITIC's finding that 97\% of large enterprises report hourly downtime costs exceeding \$100{,}000~\citep{itic2026downtime}. In both cases, we use these numbers directly, i.e., we assume a prompt response mitigating the issue within one hour. For the high tier, we ground the estimate in the medical device industry: McKinsey estimates that significant quality events (recalls, FDA enforcement actions, and litigation) cost the global medical device industry \$7--8.5 billion annually~\citep{mckinsey2026medtech}, and software defects caused an estimated 8.2\% of the 1{,}059 US medical device recall events recorded in 2024, i.e., about 87 events~\citep{meddeviceguide2026recalls}. Distributing the midpoint of McKinsey's estimate (\$7.75 billion) proportionally over the 1{,}059 recall events gives an average cost of about \$7 million per event, which we use as our high-tier estimate.

For security bugs, i.e., bugs that are exploited after reaching production, we ground our estimate in reported breach and incident costs from CISA's systematic review and cross-validation of cyber incident cost data~\citep{cisa2020cybercost}, which pools several commercial claims datasets and recommends using the median rather than the mean per-incident cost, since the underlying loss distributions are heavily right-skewed. For the low tier, we use the median total incident cost of \$48{,}000 reported for small and mid-sized businesses in NetDiligence's 2019 Cyber Claims Study (2{,}003 claims, revenue-based segmentation), as tabulated in the CISA review~\citep{cisa2020cybercost}. We consider this representative of the operator of a small web application. For the medium tier, we use the median total incident cost of \$1.0 million reported for large entities in the same NetDiligence (2019) dataset (78 claims)~\citep{cisa2020cybercost}, which reflects the larger, more heavily regulated organizations that make up a typical enterprise. For the high tier, we use the cost of the 2017 WannaCry ransomware attack on the UK's National Health Service, which exploited a security vulnerability to disrupt hospital operations and cancel 19{,}000 patient appointments at an estimated cost of \pounds92 million (about \$120 million), as reported by the UK Department of Health and Social Care~\citep{nhe2018wannacry}. We use this figure because the CISA review contains no comparable estimate for safety-critical or healthcare systems, and because it links a real exploit to the disruption of safety-critical services rather than only to data loss.

The resulting expected costs per escaped bug are reported in absolute dollars in Table~\ref{tab:scenario:bugimpact}. They increase substantially from the low to the high tier, because the same kind of bug causes far higher costs in systems with larger user bases, stricter regulatory obligations, or direct safety implications.

\begin{table}[t]
\caption{Bug impact parameters across scenarios. Low represents a simple web application, medium an enterprise system, and high a safety-critical system.}
\label{tab:scenario:bugimpact}
\footnotesize
\begin{tabularx}{\linewidth}{p{1.8cm}rrrX}
\toprule
Symbol & Low & Medium & High & Source \\
\midrule
$\omega_\mathrm{min}$ &      100 &      150 &      250 & 2h of engineering time to repair after release \\[2pt]
$\omega_\mathrm{maj}$ & 25{,}000 & 300{,}000 & 7{,}000{,}000 & Downtime cost estimates~\citep{itic2026downtime,cloudqa2026bugcost,mckinsey2026medtech,meddeviceguide2026recalls} \\[2pt]
$\omega_\mathrm{sec}$ & 48{,}000 & 1{,}000{,}000 & 120{,}000{,}000 & Breach/incident cost estimates~\citep{cisa2020cybercost,nhe2018wannacry} \\
\bottomrule
\end{tabularx}
\end{table}

% ---- 4.3 ----
\subsection{Experiment configurations}
\label{sec:setup:sweep}

\paragraph{Scenario grid.}
Since we have five orthogonal aspects, three of them with three levels and both the project dynamics and the agent costs with four, we have a total of $3^3 \cdot 4 \cdot 4 = 432$ scenarios. We solve each of these problems with the evolutionary search of Section~\ref{sec:model:search}, which yields 432 searches in total.
All scenarios share the same initial state, simulation horizon, and Monte Carlo protocol, so that any difference between two results is attributable to the scenario parameters and not to the experimental setup. The constants are summarized in the upper part of Table~\ref{tab:setup:experiment}. We simulate a green-field project, i.e., every component of the initial state $S(0)$ is empty, meaning no feature backlog, no bugs in any of the twelve bug queues, no technical debt, empty human capacity queues, no tests, and no developer documentation. We use short one-week sprints and two-year projects. We chose a long time horizon so that debt can accumulate and its negative effects become observable. We use $K' = 100$ replications per configuration during the search and $K = 1{,}000$ replications for the re-simulation of the indistinguishable set. 

The lower part of Table~\ref{tab:setup:experiment} gives the search range in the sense of Section~\ref{sec:model:search}. The staffing levels range from a quarter-FTE up to ten FTE, i.e., a ``two-pizza team''. The range extends below one FTE, because at the low project tier only $r_\mathrm{arrival} = 3$ features arrive per sprint, which a single full-time developer supported by agents exceeds by a wide margin. Part-time staffing is therefore relevant for such projects. Even at the high project tier, a quarter FTE supported by agents can deliver the required throughput. Larger teams are only chosen when they reduce the costs. The range also suffices for the very high project tier, even though the arrival speed there is far beyond what such a team could implement itself. However, the throughput saturation bounds of Eq.~\eqref{eq:sat_bound} limit a composition to 37 supervised agents at that tier, which three FTE can already monitor (Eq.~\eqref{eq:feas_monitoring}). Hence, the required throughput remains achievable.

The human-capacity condition of Eq.~\eqref{eq:feas_monitoring} limits the number of supervised agents per staffing level: with $H_\mathrm{mon,overhead} = 2.5$\,h and the documentation reduction factors $\Gamma_\mathrm{doc} \in \{0, 0.34, 0.40\}$ of Table~\ref{tab:scenario:debt}, monitoring alone exhausts a full-time developer's net capacity at roughly 16, 24, and 27 supervised agents \emph{per FTE}, respectively.

The release cadences $\mathcal{R} = \{1, 2, 4, 13\}$ cover continuous delivery (a release after every sprint), a biweekly, a monthly, and a quarterly release train, and the feature-freeze lengths $\mathcal{F}_\mathrm{freeze} = \{0, 1, 2\}$ cover no freeze, a one-week, and a two-week freeze. Both are combined subject to $F_\mathrm{freeze} \leq R - 1$ as described in Section~\ref{sec:model:search}, so $R = 1$ admits only $F_\mathrm{freeze} = 0$. The developers' bug-fixing share $f_\mathrm{fix}$ ranges from pure feature work to pure bug fixing in steps of 10\%. We set the tolerance of the throughput constraint of Eq.~\eqref{eq:throughput_constraint} to $\epsilon = 0.02$, i.e., require delivery of at least 98\% of the arriving features. 

The evolutionary search receives a budget of $E = 20{,}000$ candidate evaluations per scenario, of which we spend 10\% on the initial pool and 90\% on the generation loop, i.e., $n_\mathrm{init} = 2{,}000$, $G = 180$ generations, and $n_\mathrm{off} = 100$ offspring per generation, with an elite of $n_\mathrm{elite} = 100$ configurations. The re-simulation of the indistinguishable set with $K$ replications is not part of this budget and happens afterwards.

\begin{table}[t]
\caption{Experiment constants and search range.}
\label{tab:setup:experiment}
\footnotesize
\begin{tabularx}{\linewidth}{p{2.0cm}>{\raggedleft\arraybackslash}p{4.2cm}X}
\toprule
Symbol & \multicolumn{1}{c}{Value} & Rationale \\
\midrule
\multicolumn{3}{l}{\emph{Monte Carlo protocol and horizon}} \\[2pt]
$S(0)$                          & $\mathbf{0}$        & Green-field project without legacy state \\[2pt]
$T$                             & 104                  & Two calendar years at weekly sprints \\[2pt]
$K'$                            & 100                 & Replications during the search \\[2pt]
$K$                             & 1{,}000             & Replications for the re-simulation of $\mathcal{I}^\star$ \\[2pt]
$B$                             & 1{,}000             & Bootstrap resamples for the 95\% CIs (Section~\ref{sec:model:mc}) \\[2pt]
$s_0$                           & 42            & Master seed \\[2pt]
\midrule
\multicolumn{3}{l}{\emph{Search range}} \\[2pt]
$\mathcal{N}$                   & $\{0.25, 0.50, 0.75, 1.00, 1.50, 2.00,$ \newline $3.00, 4.00, 5.00, 7.00, 10.00\}$ & Quarter-FTE to a two-pizza team \\[2pt]
$\mathcal{R}$                   & $\{1,2,4,13\}$      & Continuous, biweekly, monthly, quarterly releases \\[2pt]
$\mathcal{F}_\mathrm{freeze}$   & $\{0,1,2\}$         & No, one-week, or two-week freeze, at most $R-1$ \\[2pt]
$\mathcal{F}_\mathrm{fix}$      & $\{0, 0.1, \ldots, 1\}$ & Share of coding time spent on bug fixing \\[2pt]
$\mathcal{F}_\mathrm{qa}$       & $\{0, 0.1, \ldots, 0.9\}$ & Share of capacity reserved for QA, human-only baseline only \\[2pt]
$\sigma$                        & 2                   & Allows up to twice the agents required for their steady-state workload \\[2pt]
$\epsilon$                      & 0.02                & At least 98\% of the arriving features must be delivered \\[2pt]
\midrule
\multicolumn{3}{l}{\emph{Search effort per scenario}} \\[2pt]
$E$                             & 20{,}000             & Candidate evaluations per scenario \\[2pt]
$n_\mathrm{init}$               & 2{,}000             & 10\% of $E$ on the initial pool \\[2pt]
$G$ / $n_\mathrm{off}$          & 180 / 100           & 90\% of $E$ on the generation loop \\[2pt]
$n_\mathrm{elite}$              & 100                 & Size of the elite that parents are drawn from \\
\bottomrule
\end{tabularx}
\end{table}

For the simulation of the human baseline, we only need to consider $N$, $R$, $F_\mathrm{freeze}$, $f_\mathrm{fix}$, and $f_\mathrm{qa}$. With the ranges of Table~\ref{tab:setup:experiment} and the constraint $F_\mathrm{freeze} \leq R-1$, these are 10{,}890 configurations in total. This is still small enough to enumerate exhaustively and we evaluate all 10{,}890 configurations with $K'$ replications, followed by the same two-stage selection of Section~\ref{sec:model:selection}, to determine the optimal teams. We run these configurations for all combinations of the human developer parameters, project dynamics, and bug costs, since agent performance and debt dynamics have no influence on this baseline. Where no configuration satisfies the throughput constraint of Eq.~\eqref{eq:throughput_constraint} (e.g., for the very high scenario), we report the least-violating one (i.e., highest throughput), mark it as inadmissible, and report how much throughput was actually achieved to put the result into perspective.

\paragraph{Use of generative AI.} Generative AI was used to generate code, as well as to assist in drafting sections of this paper. All design decisions, parameter choices, and interpretations of the results were made by the author. All output from the generative AI was carefully reviewed and verified for accuracy and relevance.

% ============================================================
\section{Results}
\label{sec:results}
% ============================================================

We now present the results of the experiments. We first analyze whether the searches converged and how precise the reported costs are (Section~\ref{sec:results:convergence}). Then, we analyze how the scenario groups affect the total costs of the optimal teams and how these costs are distributed between the three cost drivers (Section~\ref{sec:results:tradeoffs}). Afterwards, we analyze the composition of the optimal results, i.e., the staffing level and the agent composition, as well as the release process (Section~\ref{sec:results:composition}). Finally, we compare the optimal mixed teams with the optimal human-only teams (Section~\ref{sec:results:baseline}).

% ---- 5.1 ----
\subsection{Search convergence}
\label{sec:results:convergence}

All 432 searches are admissible, i.e., they found at least one configuration that satisfies the throughput constraint. For the human-only baseline, 9 of the 36 enumerations are inadmissible, namely those with very high velocity. This is not surprising, since the desired throughput of 1{,}000 issues per sprint is unrealistic with a single team and would require a complex multi-team approach within a human-only setting: the best human-only team of only senior developers manages to deliver 310 features per sprint.

\begin{figure}[t]
  \centering
  \includegraphics[width=\linewidth]{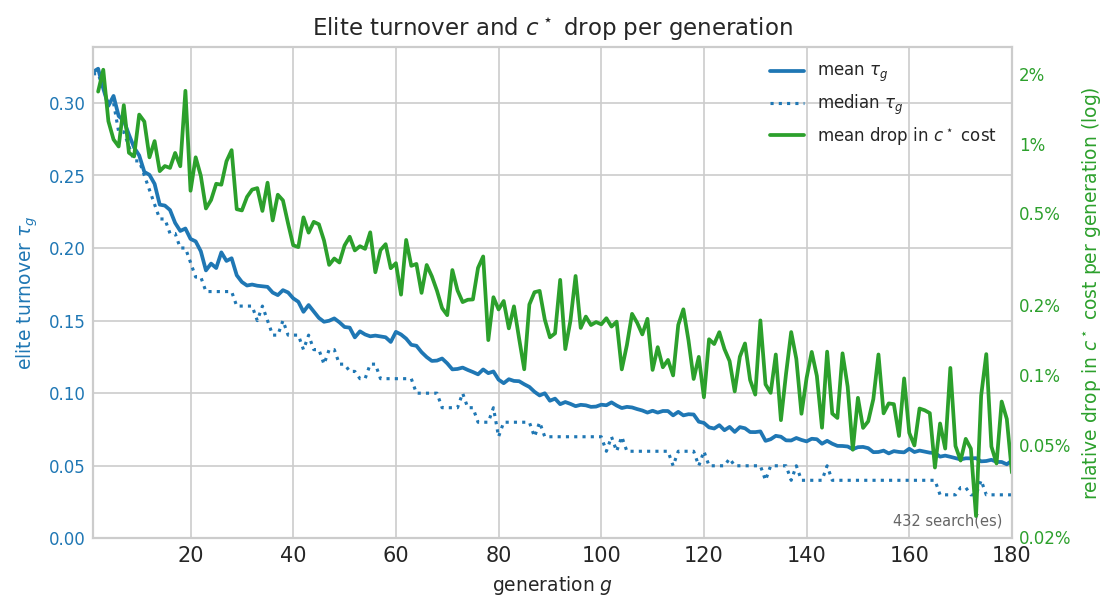}
  \caption{Elite turnover $\tau_g$ (left y-axis) and the mean absolute relative change of $c^\star$ per generation (right y-axis).}
  \label{fig:results:turnover}
\end{figure}

Figure~\ref{fig:results:turnover} shows the convergence behavior of the evolutionary search. The figure shows two aspects of the search. The blue line shows the \emph{elite turnover} defined as
\begin{equation}
  \tau_g \;=\; \frac{\bigl|\mathcal{E}_g \setminus \mathcal{E}_{g-1}\bigr|}{n_\mathrm{elite}},
  \qquad g = 1, \ldots, G,
  \label{eq:elite_turnover}
\end{equation}
i.e., the share of the elite $\mathcal{E}_g$ of generation $g$ that was not already in the elite of the preceding generation. The green line shows the mean absolute relative change of the best found result $c^\star_g$ (hereafter referred to as the incumbent) from one generation to the next, i.e., 
\begin{equation}
  \Delta c^\star_g = |c^\star_g - c^\star_{g-1}| / c^\star_{g-1}.
  \label{eq:incumbent_change}
\end{equation}
Both measures are aggregated over all 432 searches and we show the mean and median of the elite turnover, and the incumbent change as mean only, since its median is exactly zero in every generation, meaning that at least half of the searches do not change the incumbent at all in each generation.

The turnover decreases steeply over the first twenty generations and continues to decrease slowly afterwards, from a mean of $\tau_1 = 0.32$ to $\tau_{180} = 0.05$. The change in $c^\star_g$ is also decreasing and is only at a mean of 0.04\% at the end of the search. Given the huge search space and the stochastic nature of the simulation, some residual turnover is not surprising and we conclude that the search has converged sufficiently to support the analysis of the results.

The second important aspect of the search convergence follows from our decision to determine the indistinguishable set $\mathcal{I}^\star$, i.e., not only the best solution $\mathbf{c}^\star$, but also all project configurations whose CI overlaps with this optimum. Table~\ref{tab:results:topset} shows how this indistinguishable set shrinks over the four stages of the selection process described in Section~\ref{sec:model:selection}. Initially, we have a very large indistinguishable set with a median of 904 project configurations. This is not surprising, since our initial number of $K'=100$ replications is relatively small, which means the CIs are wide. Once the parsimony tie-break removes configurations for which a smaller and cheaper team exists, the indistinguishable set shrinks to a median of 324 configurations. The re-simulation with $K=1{,}000$ replications further reduces the indistinguishable set to a median of 75.5 configurations by shrinking the CIs. Finally, the second tie-break reduces the indistinguishable set to a median of 62 configurations, i.e., there is no single best team configuration but rather a space of options that are optimal when accounting for the inherent uncertainty of the development process. For the human-only baseline, there are far fewer indistinguishable configurations, with a median of 3 configurations after the first tie-break and 2 configurations after the re-simulation at $K$. This is expected, because the human-only baseline has far fewer degrees of freedom than the agentic teams, e.g., it is not possible to trade off different agent types for quality assurance against each other. 

\begin{table}[t]
\caption{Size of the indistinguishable set $\mathcal{I}^\star$ at the four stages of Section~\ref{sec:model:selection}, as median [interquartile range] and maximum over the searches that ended admissible.}
\label{tab:results:topset}
\footnotesize
\begin{tabularx}{\linewidth}{Xrrrr}
\toprule
 & \multicolumn{2}{c}{Search (432)} & \multicolumn{2}{c}{Human-only (27)} \\
\cmidrule(lr){2-3}\cmidrule(lr){4-5}
Stage & Median [IQR] & Max & Median [IQR] & Max \\
\midrule
Preliminary, at $K'$              & 904 $[455, 1{,}390]$ & 2{,}710 & 3 $[1, 12]$ & 140 \\[2pt]
After the parsimony tie-break     & 324 $[174, 476]$ & 1{,}210 & 3 $[1, 12]$ & 140 \\[2pt]
After the re-simulation at $K$    & 75.5 $[33, 127]$ & 416      & 2 $[1, 4]$  & 59 \\[2pt]
After the second tie-break        & 62 $[26, 103]$   & 345      & 2 $[1, 4]$  & 59 \\
\bottomrule
\end{tabularx}
\end{table}

Given our Monte Carlo budget of $K=1{,}000$ replications, the reported costs are also sufficiently precise.
Averaged over the final $\mathcal{I}^\star$ of a search, the relative half-width of the confidence interval
\begin{equation}
  \rho_\mathbf{c} \;=\; \frac{C^+_\mathbf{c} - C^-_\mathbf{c}}{2\,C_\mathbf{c}}
  \label{eq:rel_halfwidth}
\end{equation}
has a median of 1.6\% $[0.8\%, 4.6\%]$. Thus, the reported costs are precise to about two percent for a typical scenario.

\begin{summarybox}
The evolutionary search converged in all 432 scenarios. At the end of the search, the elite turnover fell to 5\% and the incumbent cost changed by only 0.04\% per generation. The reported costs are precise to about two percent.
\end{summarybox}

\subsection{Cost drivers}
\label{sec:results:tradeoffs}

Before we look at the structure of the optimal teams, we further validate our simulation model empirically by analyzing the costs of the optimal configurations $\mathbf{c}^\star$ and how they are affected by the scenario groups: if we observed unexpected patterns (e.g., more expensive bugs leading to lower total costs), this would indicate a problem with the model. 

We use a simple linear regression model on the logarithmic costs with a one-hot encoding of the scenario groups to quantify the effect of each group on the total costs, i.e.,
\begin{equation}
  \log_{10} C_\mathbf{c} \;=\; \beta_0 + \sum_{g=1}^{5} \sum_{\ell \neq \ell^0_g} \beta_{g,\ell} \, I_{g,\ell},
  \label{eq:linreg}
\end{equation}
where $g$ indexes the five scenario groups, $\ell$ the tiers of group $g$, $\ell^0_g$ is its cheapest tier, which serves as the reference, and $I_{g,\ell}$ is the indicator of tier $\ell$ of group $g$. This model achieves a very good fit with $R^2 = 0.938$, i.e., the scenario groups explain 93.8\% of the variance of the log cost across the grid. This shows that the impact of the scenario groups on the costs is largely independent, which is expected since the groups are orthogonal by design. This allows us to analyze the effect of each group in isolation.

\begin{figure}[t]
  \centering
  \includegraphics[width=\linewidth]{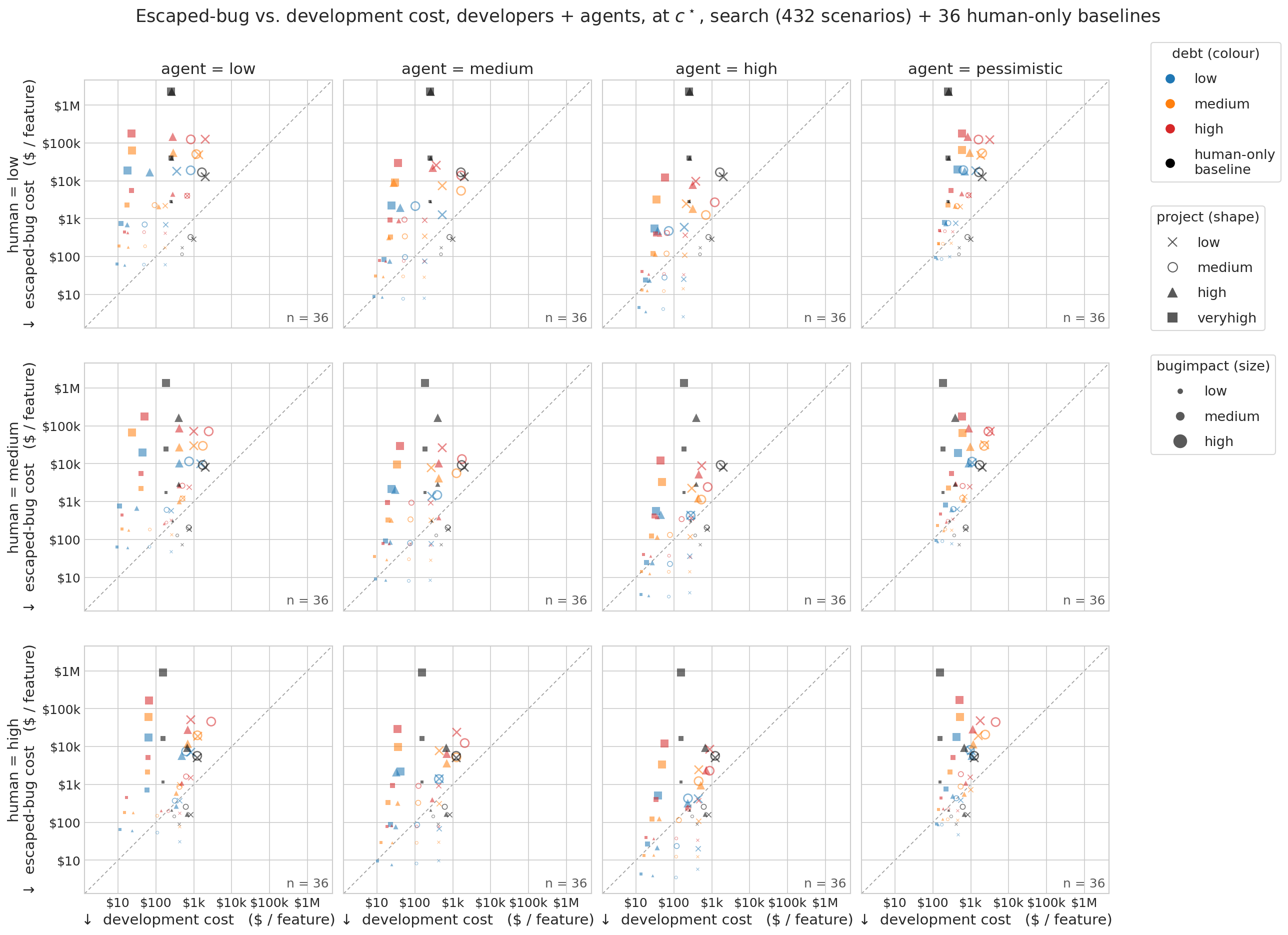}
  \caption{Costs of bugs that escaped to releases versus the costs of feature implementation, along all five scenario groups.}
  \label{fig:results:qualitydev}
\end{figure}

The coefficients of the model are shown in Table~\ref{tab:results:multipliers} relative to the cheapest tier of each group, e.g., the high bug impact tier is 72.6 times more expensive than the low bug impact tier. Bug impact is also the largest driver of the total costs, i.e., projects for which escaped bugs are expensive are also most expensive overall. Figure~\ref{fig:results:qualitydev} further highlights this, as the larger markers for the more expensive bug tiers shift the costs to the upper-left side of the plot. Even at the low tier, the costs for the escaped bugs are close to the diagonal, i.e., still about equal to the costs of the team for the feature implementation. 

The next insight is that for both humans and agents, the cheapest tier is not the low tier with the lowest personnel or token costs, respectively. For the human developers, the difference between the tiers is small, i.e., the relationship between salaries and capability seems to trade off well. For agents, the differences are more pronounced: the increase in capability of the high tier is apparently worth the increase in token costs, i.e., the overall costs increase by a factor of 7.6 at the low tier, even though the tokens are cheaper. As expected, the pessimistic tier is most expensive, with nearly ten times the costs of the high tier. Figure~\ref{fig:results:qualitydev} and Figure~\ref{fig:results:humanagent} show this more concretely: while the agent costs per feature increase with the agent tiers, the costs due to escaped bugs drop more strongly, leading to the observed reduction in overall costs with more expensive agents. 

The debt and project dynamics also impact the costs in the expected directions: debt dynamics increase the costs and the simulation shows that this can be up to a factor of 4.8. Figure~\ref{fig:results:qualitydev} shows that higher debt-dynamic tiers mainly increase the number of escaped bugs, which is the expected relation between internal quality (debt) and external quality (escaped bugs). 

The project dynamics have a smaller effect, with a factor of 2.0 between the low and the high tier. Figure~\ref{fig:results:qualitydev} shows that this is due to the lower costs per feature, while the costs for bugs remain unchanged, indicating that the more dynamic projects with more features allow for a better allocation of resources. However, as the very high fourth tier shows, this effect is not monotonic and the very high tier has 1.3 times the costs of the high tier. A possible reason for this is that due to the high speed, there are more outlier releases in which more bugs are produced, which then cannot all be caught and fixed in time. 

Finally, Figure~\ref{fig:results:qualitydev} also shows that the impact on the costs for the human-only baseline follows the same patterns as for the agentic teams.

\begin{summarybox}
The scenario groups affect simulated costs in expected ways, which validates the simulation model, with the cost of escaped bugs as the dominant driver for the total costs. The simulation shows that high capability can lead to lower overall costs, even if tokens and salaries are more expensive, due to the higher quality of the outputs.
\end{summarybox}

\begin{table}[t]
\caption{Multiplicative effect of each scenario group on the cost per feature of the optimal team, relative to the cheapest tier of that group.}
\label{tab:results:multipliers}
\footnotesize
\begin{tabularx}{\linewidth}{Xrrrr}
\toprule
Group & Low & Medium & High & Fourth tier \\
\midrule
Bug impact         & 1.0 & 4.6 & \textbf{72.6}  & --- \\[2pt]
Agent costs and throughput & 7.6 & 1.9 & 1.0            & 9.5 (pess.) \\[2pt]
Debt dynamics      & 1.0 & 2.3 & 4.8            & --- \\[2pt]
Project dynamics   & 2.0 & 1.3 & 1.0            & 1.3 (v.\ high) \\[2pt]
Human developers   & 1.1 & 1.0 & 1.0            & --- \\
\bottomrule
\end{tabularx}
\end{table}

% ---- 5.3 ----
\subsection{Composition of costs}
\label{sec:results:composition}

We now analyze the structure of the costs of the optimal results. We start with the release process, since this affects how regularly bugs may escape to production and if there is a dedicated freeze period to prevent this. Then, we analyze how costs split between investment in human developer capacity and tokens for agents. Finally, we look at the composition of the agentic teams, i.e., how token budgets are distributed between different agent roles. For simplicity, we only report the results for the single best solution $\mathbf{c}^\star$. Our online materials contain additional results with the mean over the indistinguishable set. The mean staffing level over $\mathcal{I}^\star$ deviates from that of $\mathbf{c}^\star$ by a median of 2\% and at most 14\%. Hence, $\mathbf{c}^\star$ is representative of the overall cost allocation.

\begin{figure}[tp]
  \centering
  \includegraphics[width=\linewidth]{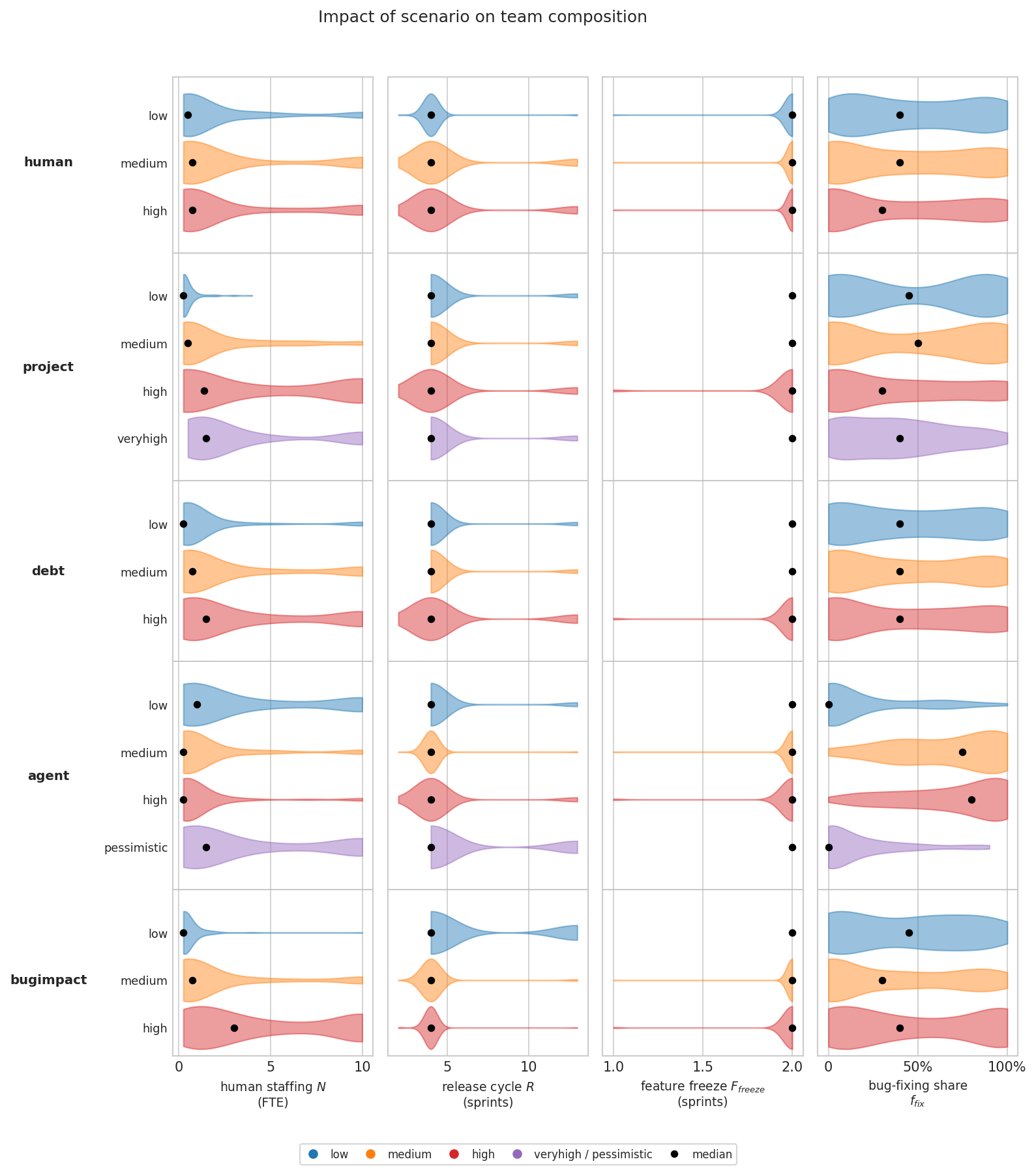}
  \caption{Impact of the scenario groups on the process controls of the optimal configurations $\mathbf{c}^\star$, with the median as a black dot.}
  \label{fig:results:drivers}
\end{figure}

The data for the release cycles is clear (Figure~\ref{fig:results:drivers}): 88.9\% of the optimal results use $R=4$ and 99.1\% use $F_\mathrm{freeze} = 2$, i.e., the longest freeze that is possible in the shortest release cycle that permits it. Interestingly, this is similar to fast release cycles in practice, as, e.g., publicly reported by \citet{gitlab2026monthlyreleases}. The likely reason for this within our simulation model is to prevent the high costs of bugs escaping to production. The two-sprint freeze is a good buffer to find and fix previously missed bugs.

\begin{figure}[t]
  \centering
  \includegraphics[width=\linewidth]{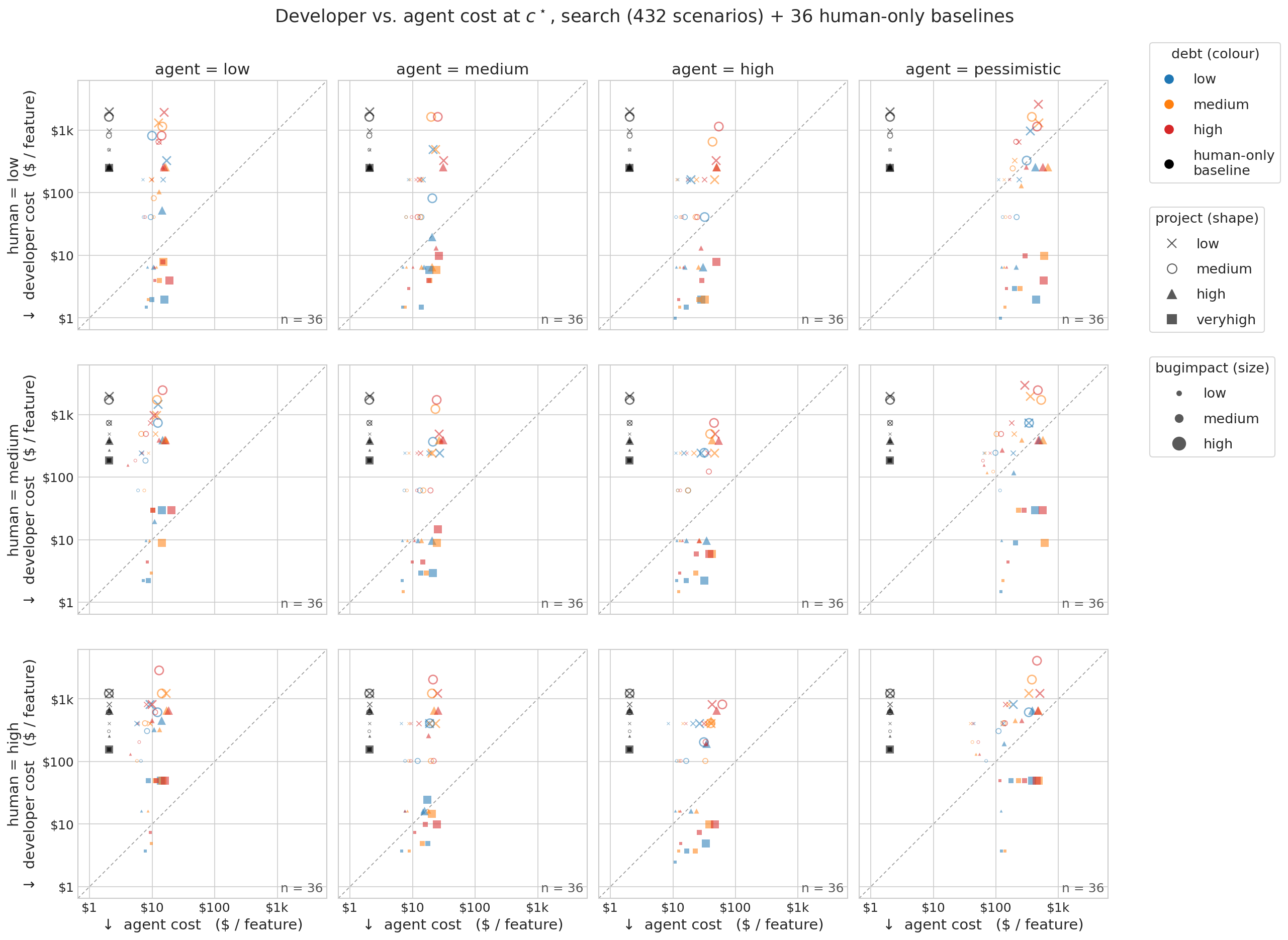}
  \caption{Developer costs versus agent costs of the optimal configurations $\mathbf{c}^\star$, along all five scenario groups.}
  \label{fig:results:humanagent}
\end{figure}

\begin{figure}[tp]
  \centering
  \includegraphics[width=\linewidth]{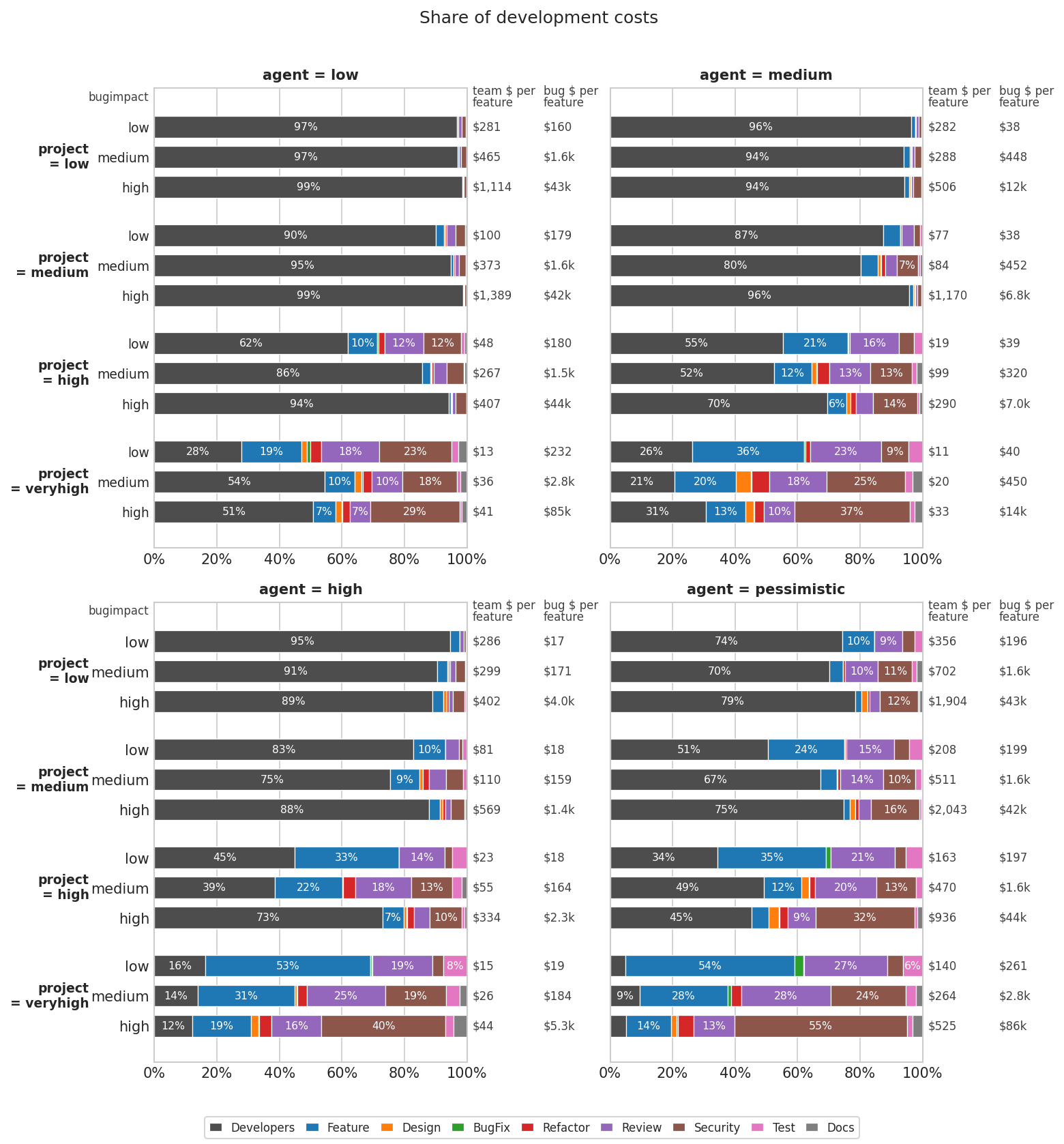}
  \caption{Distribution of the developer and agent costs of the optimal configurations $\mathbf{c}^\star$.}
  \label{fig:results:teambill}
\end{figure}

Figure~\ref{fig:results:humanagent} shows the developer costs against the agent costs and Figure~\ref{fig:results:teambill} shows a more fine-grained picture that includes how the costs are distributed between different agent roles. The developer costs are higher for 65\% of the optimal teams, with a median ratio of agent to developer costs of $0.32$. The project velocity has the largest effect on this ratio: higher velocity shifts spending from developers to agents, which are needed to handle the demand. Figure~\ref{fig:results:drivers} shows that while the staffing level does increase with project velocity, this increase is modest. The staffing level depends more strongly on the agent quality (minimal staffing for the medium and high-quality tiers), bug impact (higher staffing to prevent expensive bugs), and debt dynamics (to make the project more robust against the negative effects of debt).

The costs of agents mainly exceed the costs of human developers at the high and very high project tiers. Even at these tiers this is not always the case, because the ratio also depends on the agent tier, the second most important driver. As we already observed in Section~\ref{sec:results:tradeoffs}, teams with high-tier agents are cheaper overall and therefore invest more in agents than teams with low- or medium-tier agents. The pessimistic agents are expensive, but are nevertheless used and shift costs towards spending for agents. How much spending shifts towards pessimistic agents depends on the human tier. With low-capability developers, relying on agents is still worthwhile, and at high velocities partly necessary. In some very high velocity scenarios, the agent spending exceeds the developer spending by a factor of over 100. For high-capability developers, this is reduced and spending for pessimistic agents is only above spending for human developers at the very high velocity tier, and even there it only reaches 37 times the costs for human developers. 

How the developers split their coding time between implementing features and fixing bugs depends almost exclusively on the agent capability (panel $f_\mathrm{fix}$ in Figure~\ref{fig:results:drivers}). With the medium and high agent tiers, the developers spend a median of 75\% and 80\% of their coding time in the development sprints on bug fixing, with the low and pessimistic tiers 0\%. These numbers need to be considered together with the release cycle of four sprints of which two are feature freezes, during which all time is invested in bug fixing. Thus, for the medium and high agent tiers, developers typically spend a total of 88--90\% of their coding time on bug fixing, whereas they split their time equally for the low and pessimistic agent tiers, investing 50\% of effort into implementing features (100\% during development sprints) and 50\% into bug fixing (100\% during freeze sprints). 

\begin{figure}[tp]
  \centering
  \includegraphics[width=\linewidth]{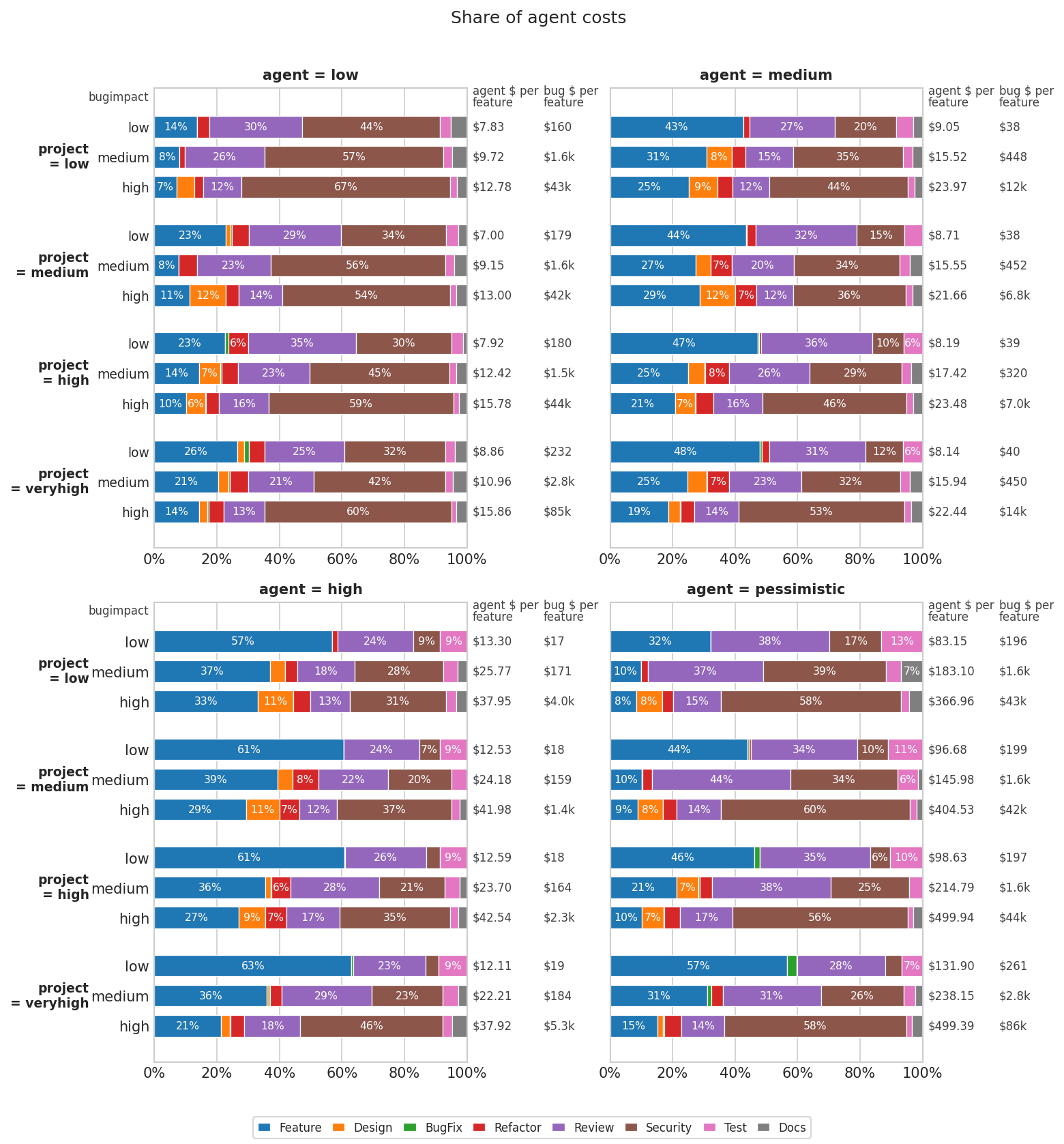}
  \caption{Distribution of the agent costs of the optimal configurations $\mathbf{c}^\star$ between the eight agent roles.}
  \label{fig:results:agentbill}
\end{figure}

Figure~\ref{fig:results:agentbill} shows how the agent costs are distributed between the eight roles. Most of the agent budget goes to quality assurance: on average, 61\% of the agent costs are spent on the Security Audit (33\%), Code Review (24\%), and Test Development Agents (4\%). The share of the Security Audit Agents is determined by the bug impact and the agent tier, i.e., the more expensive bugs are and the less capable agents are at finding them, the more spending is allocated to prevent the most critical bugs. The second largest share goes to the Feature Implementation Agents (28\%). The concrete values vary strongly and low-capability agents have a smaller implementation share than high-capability agents. This indicates that the model prefers the less bug-prone human development, which is consistent with the developers' split between feature implementation and bug fixing. The share of implementation agents in the overall costs for agents does not seem to depend on the project dynamics, i.e., even for the very high project tier, the agentic feature implementation generally behaves the same as for projects with only few features per sprint. 

The relevance of the three debt-related agents for Design, Refactoring, and Documentation depends on the impact of bugs: if bugs are cheap, these agents are often not used at all, or only with a very small overall share of the budget. When bugs are expensive, spending on debt reduction becomes worthwhile. Debt affects code quality directly, because it raises the bug injection rate, and indirectly, because it lowers the detection rate of reviews. The data also shows that there are relatively few Bug Fixing Agents. According to the simulation, bug fixing should therefore remain predominantly a human task, as this seems to be more cost-efficient under our assumptions. 

\begin{summarybox}
The optimal teams release every four sprints with a two-sprint feature freeze. At realistic project velocities they spend more on developers than on agents, and higher velocity and more capable agents shift the spending towards agents. Most of the agent budget goes to quality assurance, above all Security Audit Agents, whose share grows with the cost of escaped bugs. Bug fixing stays mainly a human task: developers spend most of their coding time on it when the agents are capable.
\end{summarybox}

% ---- 5.4 ----
\subsection{Comparison with human baseline}
\label{sec:results:baseline}

Now that we understand how the teams are structured and how the costs are distributed, we can further analyze whether the agents actually reduce the costs of software development, using our human-only baseline as reference. Since the baseline does not depend on the debt dynamics and the agent tiers, we compare each baseline to the 12 optimal teams with the same human, project, and bug impact tiers. Table~\ref{tab:results:baseline} shows the ratio between the costs of the baseline and of the optimal team, i.e., a ratio above one means that the team with agents is cheaper. We exclude the very high project tier from this comparison, because the human-only teams cannot achieve the required velocity (Section~\ref{sec:results:convergence}). 

Overall, the teams with agents are cheaper in 71.6\% of the scenarios, with a median cost ratio of 1.93, i.e., almost half of the costs are saved. However, this median hides the strong effect of the agent tier and the project dynamics on the cost savings: the high agent tier almost always saves costs (98.8\% of the simulated scenarios), while the low and pessimistic tiers are closer to a coin flip (55.6\% and 44.4\% of the scenarios, respectively). For all tiers, the cost savings increase with the project dynamics, i.e., the more features per sprint, the more the agents help to reduce costs. The cost drivers identified above apply here as well. Expensive bugs amplify the disadvantage of low-quality agents, and high debt dynamics shift the balance towards human developers.

\begin{table}[t]
\caption{Cost ratio between the human-only baseline and the optimal team with agents, as median per project tier and overall median [interquartile range].}
\label{tab:results:baseline}
\footnotesize
\begin{tabularx}{\linewidth}{Xrrrrr}
\toprule
 & \multicolumn{3}{c}{Project dynamics} & & Agents \\
\cmidrule(lr){2-4}
Agent tier & Low & Medium & High & Overall [IQR] & cheaper \\
\midrule
Low         & 0.81 & 0.89 & 4.70  & 1.11 $[0.52, 2.31]$  & 55.6\% \\[2pt]
Medium      & 1.71 & 3.20 & 15.08 & 3.35 $[1.38, 8.18]$  & 87.7\% \\[2pt]
High        & 2.18 & 5.66 & 23.79 & 6.13 $[2.40, 15.03]$ & 98.8\% \\[2pt]
Pessimistic & 0.74 & 0.88 & 2.10  & 0.93 $[0.45, 1.63]$  & 44.4\% \\
\midrule
All         & 1.12 & 1.74 & 8.65  & 1.93 $[0.88, 6.50]$  & 71.6\% \\
\bottomrule
\end{tabularx}
\end{table}

\begin{summarybox}
Teams with medium- and high-tier agents are cheaper than human-only teams in almost all scenarios, while low-tier and pessimistic agents offer no consistent advantage. The advantage of agents grows with the project velocity.
\end{summarybox}

% ============================================================
\section{Discussion}
\label{sec:discussion}
% ============================================================

We now discuss what our results mean for the research question and what they imply for practice.

\subsection{Answer to the research question}
\label{sec:discussion:rq}

As in traditional software development, there is no single optimal team configuration for AI-native development. How costs should be distributed between human developers and agents and for which tasks money should be spent on agents depends on the project. The two key factors that increase the spending on agents are the costs associated with bugs and the overall project velocity. Additionally, higher debt dynamics require better agents or stronger human involvement. When selecting agents, our results show that capability is more important than the price per token, since better agents inject fewer bugs and require less supervision, which outweighs their higher token costs.

The task distribution between the humans and agents is fairly consistent in our simulations. When agents are capable, developers spend 88--90\% of their coding time on bug fixing, while few (if any) agents are used for bug fixing. In turn, the agents implement features and, above all, detect bugs: on average, 61\% of the agent budget goes to the Security Audit, Code Review, and Test Development Agents, and 28\% to the Feature Implementation Agents. When the agents are less capable, the developers spend about half of their coding time on feature implementation and half on bug fixing.

Regarding the cost split between humans and agents, our results show that at realistic velocities, most money is spent on developers. The developer costs exceed the agent costs for 65\% of the optimal teams, with a median ratio of agent to developer costs of 0.32. At high and very high project velocities, spending shifts towards agents. Thus, the project velocity determines how much is spent on agents, while the capability of the agents determines whether agents are worth using at all.

\subsection{Revisiting the token budget}
\label{sec:discussion:huang}

We started this work with Jensen Huang's statement that developers should consume at least half of their salary in tokens. Our results show that this rule is not far off but depends strongly on project factors: if the project velocity is low, the agents are not very capable, or the bugs are expensive, then the optimal teams often spend less than half of the developer costs on tokens. If the project velocity is high, the agents are capable, and bugs are cheap, then the optimal teams spend more than half of the developer costs on tokens. 

Moreover, our results generally show that high token spending is not a sign of efficient agent use. The highest ratio of agent to developer costs occurs with the pessimistic agents. However, these teams are cheaper than the human-only baseline in less than half of the scenarios we considered. High token consumption may therefore indicate that agents are expensive relative to what they deliver. We thus believe that token consumption is a poor indicator for the productivity of AI-native development. A better measure is the total cost per delivered feature, including the cost of the bugs that reach production.

\subsection{Implications for practice}
\label{sec:discussion:practice}

Our results have several direct implications for practitioners who want to shift towards agentic software development: 
\begin{itemize}
  \item Focus on capability, not price per token, since token prices are offset by cost reductions through the higher output quality.
  \item Ensure that agents have the required capability for your use case, since our simulations show that the value of agents strongly depends on what you assume is their capability to deliver bug-free code and to detect bugs.
  \item Spend on quality assurance, since this scales well with agents and can prevent downstream costs of escaped bugs.
  \item Projects that might have been out of reach due to their velocity might be achievable with agents, since throughput can be increased by scaling both implementation and quality assurance agents.
  \item Monitor technical debt, since agents tend to increase it~\citep{he2026cursordebt}, which can lead to a long-term increase in costs if not managed properly.
\end{itemize}

% ============================================================
\section{Conclusion}
\label{sec:conclusion}
% ============================================================

We developed a simulation model of software development teams for teams of human developers and AI agents in eight different roles. Using an evolutionary search over 432 scenarios, we determined the team configurations that deliver the work a project demands at the lowest total cost per feature to analyze the cost structure and team structure of AI-native development. Our results show that there is no single optimal team for AI-native software development. The capability of the agents matters more than their price per token, because capable agents inject fewer bugs and require less supervision. Whether human-only or AI-native development is cheaper mainly depends on the specific context and the capabilities of the agents involved. Highly capable agents make development cheaper, whereas low-capability agents offer no consistent advantage. Escaped bugs are the main cost driver, which is why most of the agent budget is spent on finding bugs. At realistic project velocities, the optimal teams still spend more on developers than on agents. Only high velocities shift the spending towards agents, which also enable velocities that human-only teams cannot achieve. Consequently, token consumption alone is a poor indicator for the efficiency of agent use, and the total cost per delivered feature is a better measure.

Future work should validate our model against data from real projects that use agents, in particular regarding the capability of agents to detect bugs and the costs of escaped bugs, which our results depend on most strongly. Moreover, the model could be extended to assess additional factors, e.g., accumulation of intent and cognitive debt~\citep{Storey2026} or to relax the simplifications discussed in Section~\ref{sec:model:limitations}. Finally, since agent capabilities and token prices change rapidly, our simulation framework can be used to re-evaluate the optimal team compositions whenever new estimates become available.

%\begin{acknowledgements}
% None.
%\end{acknowledgements}

\section*{Declarations}

\paragraph{Funding.} No funding was received for conducting this study.

\paragraph{Competing interests.} The author has no competing interests to declare that are relevant to the content of this article.

\paragraph{Ethics approval and consent to participate.} Not applicable, since this study does not involve human participants or animals.

\paragraph{Author contributions.} Steffen Herbold is the sole author and carried out the conceptualization, methodology, software, analysis, and writing of this article.

\paragraph{Data, materials, and code availability.} All code is available on GitHub: \url{https://github.com/aieng-lab/se-with-ai-simulation}. The raw results of the experiments and the consolidated results from which all tables and figures of the results section are derived are available on Zenodo: \url{https://doi.org/10.5281/zenodo.23358924}.

% ============================================================
% APPENDICES
% ============================================================

\bibliographystyle{spbasic}      % basic style, author-year citations
\bibliography{literature}        % name your BibTeX data base

\end{document}